%% file: front.tex
\input header.tex
\input defs.tex

\input refs.tex

\hfuzz=1pt

\TITLE Scaling Variables and Stability of Hyperbolic Fronts

\AUTHOR Th. Gallay and G. Raugel

\FROM CNRS et Universit\'e de Paris-Sud
      URA 760, Analyse Num\'erique et EDP
      B\^atiment 425
      F-91405 Orsay Cedex, France
      Thierry.Gallay@math.u-psud.fr
      Genevieve.Raugel@math.u-psud.fr
\ENDTITLE

\ABSTRACT 
We consider the damped hyperbolic equation
$$
   \epsilon u_{tt} + u_t \,=\, u_{xx} + F(u)~, \quad x \in \real~, \quad
   t \ge 0~, \eqno(1)
$$
where $\epsilon$ is a positive, not necessarily small parameter. We assume
that $F(0) = F(1) = 0$ and that $F$ is concave on the interval $[0,1]$. 
Under these hypotheses, Eq.(1) has a family of monotone travelling wave
solutions (or propagating fronts) connecting the equilibria $u=0$ and 
$u=1$. This family is indexed by a parameter $c \ge c_*$ related to 
the speed of the front. In the critical case $c=c_*$, we prove that the
travelling wave is asymptotically stable with respect to perturbations
in a weighted Sobolev space. In addition, we show that the perturbations
decay to zero like $t^{-3/2}$ as $t \to +\infty$ and approach a universal
self-similar profile, which is independent of $\epsilon$, $F$ and of the
initial data. In particular, our solutions behave for large times like
those of the parabolic equation obtained by setting $\epsilon = 0$ in Eq.(1).
The proof of our results relies on careful energy estimates for the equation
(1) rewritten in self-similar variables $x/\sqrt{t}$, $\log t$. 

\medskip
\noindent Keywords : damped wave equation, travelling wave, stability,
asymptotic behavior, self-similar variables

\noindent AMS classification codes (1991) : 35B40, 35B35, 35B30, 35L05, 
35C20

\ENDABSTRACT

\SECTION Introduction

In this paper, we study the asymptotic stability of travelling wave solutions
to nonlinear damped hyperbolic equations on the real line. Besides describing
the propagation of voltage along nonlinear transmission lines, these equations
have been proposed as mathematical models for spreading and interacting 
particles \ref{DO}, \ref{Ha2}, \ref{Ha3}. In the latter context, they provide 
an alternative to the reaction-diffusion systems which are very common in 
chemistry and biology, especially in genetics and population dynamics
\ref{Mu}. The two classes of models differ by the choice of the stochastic 
process describing the spatial spread of the individuals: instead of Brownian
motion, the damped hyperbolic equations are based on a more realistic velocity
jump process which takes into account the inertia of the particles \ref{Go}, 
\ref{Kac}, \ref{Za}. Since this process is asymptotically diffusive, the 
long-time behavior of the solutions is expected to be essentially parabolic 
\ref{GR2}. 

We study here the simple case of a scalar equation with a nonlinearity of
``monostable'' type. To be specific, we consider the equation
$$
   \epsilon U_{\t\t} + U_\t \,=\, U_{\x\x} + \FF(U)~, \EQ(U)
$$
where $\x \in \real$, $\t \ge 0$, and $\epsilon$ is a positive, 
{\sl not necessarily small} parameter. We assume that the nonlinearity
$\FF \in \CC^2(\real)$ satisfies 
$$
   \FF(0) = \FF(1) = 0~, \quad \FF'(0) > 0~, \quad \FF'(1) < 0~, \quad
   \FF''(U) \le 0 \hbox{~~for~~} U \in [0,1]~. \EQ(nonlin)
$$
In particular, $U = 1$ is a stable equilibrium of Eq.\equ(U), and $U = 0$ is 
unstable. A typical nonlinearity satisfying \equ(nonlin) is $\FF(U) = U-U^m$, 
with $m \ge 2$. 

Under the assumptions \equ(nonlin), Eq.\equ(U) has monotone travelling wave 
solutions (or propagating fronts) connecting the equilibrium states 
$U=1$ and $U=0$ \ref{Ha1}, \ref{GR1}. Indeed, choosing $c > 0$ and
setting $U(\x,\t) = h(\sqrt{1+\epsilon c^2}\x-c\t)$, we obtain for $h$ 
the ordinary differential equation
$$
   h''(\xi) + ch'(\xi) + \FF(h(\xi)) \,=\, 0~, \quad \xi \in \real~.
   \EQ(front)
$$
Eq.\equ(front) is known to have a strictly decreasing solution satisfying 
$h(-\infty) = 1$ and $h(+\infty) = 0$ if and only if $c \ge c_* = 
2\sqrt{\FF'(0)}$ \ref{KPP}, \ref{AW}. This solution is unique up to 
translations in the variable $\xi$. Thus, Eq.\equ(U) has a family of 
monotone travelling waves indexed by the speed parameter $c \ge c_*$. 
Note that the actual speed of the wave is not $c$, but $c/\sqrt{1{+}\epsilon 
c^2}$, a quantity which is bounded by $1/\sqrt{\epsilon}$ for all $c \ge c_*$. 

In an earlier paper \ref{GR1}, we investigated the stability of the 
travelling waves of \equ(U) in the case where $\FF(U) = U-U^2$. 
In particular, we showed that, for all $\epsilon > 0$ and all $c \ge c_*$,
the front $h$ is asymptotically stable with respect to small
perturbations in a weighted Sobolev space (with exponential weight). This 
local stability result holds in fact for all nonlinearities satisfying 
\equ(nonlin), see \ref{GR3}. In addition, if $\epsilon > 0$ is sufficiently 
small, we proved in \ref{GR1} that the front $h$ is stable with respect to 
large perturbations, provided  some positivity conditions are fulfilled. 
This global stability property relies on the hyperbolic Maximum Principle,
and can also be extended to more general nonlinearities \ref{GR3}. Finally, 
we showed in all cases that the perturbations converge uniformly to zero 
faster than $\t^{-1/4}$ as $\t \to +\infty$. 

When $\epsilon \to 0$, Eq.\equ(U) reduces to the semilinear parabolic 
equation $U_{\t} = U_{\x\x} + \FF(U)$ which has been intensively studied
since the pioneering works of Fisher \ref{Fi} and Kolmogorov, Petrovskii and 
Piskunov \ref{KPP}. Using the parabolic Maximum Principle and probabilistic 
techniques, the convergence of a large class of solutions to travelling waves
has been established \ref{AW}, \ref{Br}. In the more general context of 
parabolic systems, a local stability analysis of the waves has been initiated
by Sattinger \ref{Sa} and extended by many authors \ref{Ki}, \ref{EW}, 
\ref{Kap}, \ref{BK1}, \ref{RK}, using resolvent estimates, energy functionals 
and renormalization techniques. In the critical case $c = c_*$, it has been 
proved by one of us \ref{Ga} that the perturbations of the front decay to 
zero like $\t^{-3/2}$ as $\t \to +\infty$ and approach a universal 
self-similar profile. The aim of the present paper is precisely to extend
this detailed convergence result to the hyperbolic case $\epsilon > 0$. 
Together with earlier results from \ref{GR1}, \ref{GR3}, this will provide a 
fairly complete picture of the stability properties of the travelling waves 
of Eq.\equ(U). 

To study the stability of the critical front $h$ with $c=c_*$, it is 
convenient to go to a moving frame using the change of variables
$U(\x,\t) = V(\sqrt{1+\epsilon c_*^2}\x - c_*\t,\t)$. The equation
for $V$ is
$$
   \epsilon V_{\t\t} + V_\t - 2\epsilon c_* V_{\xi\t} \,=\, 
   V_{\xi\xi} + c_* V_\xi + \FF(V)~, \EQ(V)
$$
where $\xi = \sqrt{1+\epsilon c_*^2}\x - c_*\t$. By construction, $h$ is 
a stationary solution of \equ(V). Following \ref{Ki}, \ref{Ga}, we consider 
perturbed solutions of the form
$$
   V(\xi,\t) \,=\, h(\xi) + w(\xi,\t) \,\equiv\, h(\xi) + 
   h'(\xi)W\left(\xi,{\t \over 1+\epsilon c_*^2}\right)~. \EQ(pert)
$$
The reason for this Ansatz is that the function $W(\xi,\tau)$ defined by 
\equ(pert) becomes asymptotically self-similar as $\t \to +\infty$, while 
the actual perturbation $w(\xi,\t)$ does not, see Corollary~1.3 below. 
Remark that $W$ is well-defined, since $h'(\xi) < 0$ for all $\xi \in \real$. 
For notational convenience, we rescale the time variable $\t$ by setting 
$\tau = \t/(1{+}\epsilon c_*^2)$. The equation for $W$ then reads
$$
   \eta W_{\tau\tau} + (1-\nu \gamma(\xi))W_\tau - 2\nu W_{\xi\tau} 
   \,=\, W_{\xi\xi} + \gamma(\xi) W_\xi + h'(\xi)W^2 
   \NN(h(\xi),h'(\xi)W)~, \EQ(W)
$$
where
$$
   \eta \,=\, {\epsilon \over (1+\epsilon c_*^2)^2}~, \quad
   \nu \,=\, {\epsilon c_* \over 1+\epsilon c_*^2}~, \quad
   \gamma(\xi) \,=\, c_* + 2 {h''(\xi) \over h'(\xi)}~, 
   \EQ(etanugam)
$$
and
$$
   \NN(a,b) \,=\, \int_0^1 (1{-}s)\FF''(a+sb)\d s \,=\, 
   {1 \over b^2} \bigl(\FF(a+b) - \FF(a) - b\FF'(a)\bigr)~. \EQ(NN)
$$

\figurewithtex 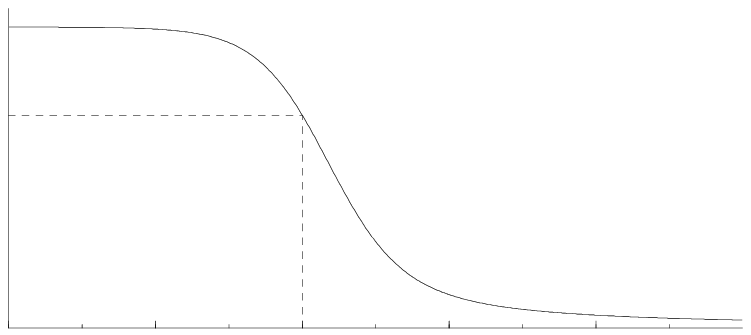 fig1.tex 4 11 The function $\gamma(\xi)$ in the  
case where $\FF(U) = U-U^2$ (hence $c_* = 2$, $\gamma_- = 2\sqrt{2}$).\cr

Before analyzing the solutions of \equ(W), we briefly comment on the 
definitions \equ(etanugam). We first remark that there is no loss
of generality in assuming $\epsilon = 1$ in Eq.\equ(U), since $(\epsilon,\FF)$
can be transformed into $(1,\epsilon\FF)$ by rescaling $\x$ and $\t$. 
However, we find more convenient to fix the nonlinearity $\FF$ and to 
consider $\epsilon$ as a free parameter. Then $c_* > 0$ is fixed, 
and $\eta,\nu$ are functions of $\epsilon$ only. These expressions are not 
independent, since $\nu^2 + \eta = \nu/c_*$. Observe also that $\eta,\nu$ 
are uniformly bounded for all $\epsilon > 0$, and converge to zero as 
$\epsilon \to 0$. 
We now list the properties of the ``drift'' $\gamma(\xi)$ which will be 
crucial for our analysis. From \ref{Sa}, \ref{AW}, we know that the front 
$h$ (with $c=c_*$) satisfies
$$
   h(\xi) \,=\, \cases{1 - a_3 \e^{\kappa\xi} + \OO(\e^{2\kappa\xi}) & as 
    $\xi \to -\infty$~, \cr
   (a_1\xi{+}a_2)\e^{-c_*\xi/2} + \OO(\xi^2\e^{-c_*\xi}) & as $\xi \to 
    +\infty$~,} \EQ(hasym)
$$
where $a_1,a_3 > 0$, $a_2 \in \real$, and $\kappa = {1 \over 2}(-c_*
+\sqrt{c_*^2-4\FF'(1)}) > 0$. Using \equ(hasym) and similar asymptotic 
expansions for the derivatives $h'$, $h''$, we obtain
$$
   \gamma(\xi) \,=\, \cases{\gamma_- + \OO(\e^{\kappa\xi}) & as $\xi \to 
   -\infty$~, \cr 2/(\xi{+}\xi_0) + \OO(\xi \e^{-c_*\xi/2}) & as $\xi \to 
   +\infty$~,} \EQ(gamasym)
$$
where $\gamma_- = c_*+2\kappa = 2\sqrt{\FF'(0)-\FF'(1)}$ and $\xi_0 = 
(a_2/a_1-2/c_*)$. It also follows from \equ(front), \equ(etanugam) that 
$$
   \gamma'(\xi) \,=\, -{1 \over 2}\gamma(\xi)^2 + 2\bigl(\FF'(0)-\FF'(h(\xi))
   \bigr)~, \quad \xi \in \real~. \EQ(gamprim)
$$
Together with \equ(nonlin), this equation implies that $-{1 \over 2}
\gamma(\xi)^2 \le \gamma'(\xi) \le 0$ for all $\xi \in \real$. Indeed, 
the lower bound on $\gamma'(\xi)$ is obvious, and the upper bound follows from
the inequality $\gamma''(\xi) + \gamma(\xi)\gamma'(\xi) \le 0$ obtained
by differentiating \equ(gamprim). In fact, we even have $\gamma'(\xi) < 0$
whenever $\gamma(\xi) < \gamma_-$. Replacing thus $h(\xi)$ by a translate, 
we may (and will always) assume that $\gamma(0) = c_*$, {\sl i.e.} 
$h''(0) = 0$, see Fig.~1. This amounts to fixing the origin in the moving 
frame.  

To study the behavior of the solutions $W$ of \equ(W), we use the 
{\sl scaling variables} or {\sl self-similar variables} defined by
$$
   x \,=\, {\xi \over \sqrt{\tau{+}\tau_0}}~, \quad t = \log(\tau{+}\tau_0)~,
   \EQ(simvar)
$$
where $\tau_0 > 0$ will be fixed later. These variables have been widely 
used to investigate the long time behavior of solutions to parabolic 
equations, in particular to prove convergence to self-similar solutions 
\ref{Kav}, \ref{EZ}, \ref{GV}, \ref{EKM}, \ref{BK2}, \ref{Wa}, \ref{GM}.
Although the scaling \equ(simvar) is parabolic in essence, we have shown in 
\ref{GR2} that self-similar variables are also a powerful tool in the realm 
of damped hyperbolic equations. The reason is that the long-time behavior of 
such systems is often determined by simpler parabolic equations, see \ref{HL}, 
\ref{Ni}, \ref{GR2} for specific examples of this phenomenon. In our case,
the result of \ref{Ga} in the parabolic limit $\epsilon = 0$ suggests that
$W(\xi,\tau)$ should behave like $\tau^{-3/2}\phi^*(\xi/\sqrt{\tau})$ as
$\tau \to +\infty$, where $\phi^*$ is given by (1.20) below. Thus, following
the method developped in \ref{GR2}, we define rescaled functions $u$ and $v$ 
by
$$
   u(x,t) \,=\, \e^{3t/2}W(x\e^{t/2},\e^t-\tau_0)~, \quad v(x,t) \,=\,
   \e^{5t/2}W_\tau(x\e^{t/2},\e^t-\tau_0)~, \EQ(sv)
$$
or equivalently
$$ \eqalign{
   W(\xi,\tau) \,&=\, {1 \over (\tau{+}\tau_0)^{3/2}} \,u\left({\xi \over
    \sqrt{\tau{+}\tau_0}}\,,\,\log(\tau{+}\tau_0)\right)~, \cr
   W_\tau(\xi,\tau) \,&=\, {1 \over (\tau{+}\tau_0)^{5/2}} \,v\left({\xi \over
    \sqrt{\tau{+}\tau_0}}\,,\,\log(\tau{+}\tau_0)\right)~.} \EQ(vs)
$$
Then the functions $u(x,t), v(x,t)$ satisfy the system
$$ \eqalign{
   &u_t - {x \over 2} u_x - {3 \over 2} u \,=\, v~, \cr
   &\eta \e^{-t}\bigl(v_t - {x \over 2}v_x - {5 \over 2}v\bigr) + 
   (1-\nu\gamma(x \e^{t/2}))v -2\nu\e^{-t/2}v_x \,=\, \cr
   & \qquad u_{xx} + \e^{t/2} \gamma(x\e^{t/2}) u_x +\e^{-t/2} h'(x\e^{t/2})
   u(x,t)^2 N(x,t)~,} \EQ(uv)
$$
where $x \in \real$, $t \ge t_0 = \log \tau_0$, and $N(x,t) = 
\NN(h(x\e^{t/2}),\e^{-3t/2}h'(x\e^{t/2})u(x,t))$. 

We next introduce the function spaces in which we shall study the
solutions of \equ(uv). For $t \ge 0$, $k \in \natural$, we denote by 
$\L^2_t$, $\H^k_t$ the weighted Lebesgue and Sobolev spaces defined by 
the norms
$$\eqalign{
  \|u\|_{\L^2_t}^2 \,&=\, \int_{-\infty}^0 \e^{2\kappa x\e^{t/2}}|u(x)|^2
   \d x + \int_0^\infty (1{+}x)^6|u(x)|^2\d x~, \cr
  \|u\|_{\H^k_t}^2 \,&=\, \sum_{i=0}^k \|\partial_x^i u\|_{\L^2_t}^2~,}
  \EQ(norms)
$$
where $\kappa$ appears in \equ(hasym). Our basic space will be the product 
$\Z_t = \H^1_t \times \L^2_t$ equipped with the standard norm 
$\|(u,v)\|_{\Z_t} = (\|u\|_{\H^1_t}^2 + \|v\|_{\L^2_t}^2)^{1/2}$. In order 
to state results which are uniform in $\epsilon$ as $\epsilon \to 0$, 
it is convenient to introduce also the quadratic form
$$
   \Phi_\eta(t,u,v) \,=\, \|u\|_{\H^1_t}^2 + \eta \e^{-t}\|v\|_{\L^2_t}^2~. 
   \EQ(Phi)
$$
{}From \equ(sv), \equ(vs), we see that $(u,v) \in \Z_t$ if and only if
$(W,W_\tau) \in \Z_0 = \H^1_0 \times \L^2_0$. Moreover, since $h',h'' = 
\OO(\e^{\kappa\xi})$ as $\xi \to -\infty$ and $h',h'' = \OO(\xi \e^{-c_*
\xi/2})$ as $\xi \to +\infty$, it is easy to verify that $(W,W_\tau) \in 
\Z_0$ if and only if the actual perturbation $w = h'W$ satisfies
$$
  \int_{-\infty}^0 (w^2 + w_\xi^2 + w_\t^2)\d \xi +
  \int_0^\infty (1+\xi)^4 \e^{c_* \xi} (w^2 + w_\xi^2 + w_\t^2)\d \xi
   \,<\, \infty~. \EQ(actual)
$$
The comparison of \equ(hasym), \equ(actual) reveals that the perturbations
we consider decay to zero slightly faster than the front $h$ itself as 
$\xi \to +\infty$. This is a necessary condition for stability, because the 
equilibrium state $U=0$ of \equ(U) is linearly unstable \ref{Sa}. In 
particular, small translations of the front $h$ are {\sl not} allowed as 
perturbations. 

Since our function space $\Z_t$ depends on time, we have to specify what 
we mean by a ``solution of \equ(uv) in $\Z_t$''. As the system \equ(uv)
has been obtained from the simpler equation \equ(W) through the change 
of variables \equ(sv), the following definition is very natural:

\global\advance\CLAIMcount by 1
\LIKEREMARK(Definition 1.1) Let $t_2 > t_1 \ge t_0$, and let $\tau_i = 
\e^{t_i}-\tau_0$ for $i=1,2$. We say that ``$(u,v) \in \CC([t_1,t_2],\Z_t)$ 
is a solution of the system \equ(uv)'' if there exists a (mild) solution 
$(W,W_\tau) \in \CC([\tau_1,\tau_2],\Z_0)$ of \equ(W) such that the relations 
\equ(sv), \equ(vs) hold. 

In particular, if $(u,v) \in \CC([t_1,t_2],\Z_t)$ is a solution of \equ(uv), 
then $(u(t),v(t)) \in \Z_t$ for all $t \in [t_1,t_2]$. However, the continuity
of $(u,v)$ with respect to $t$ has to be understood as the continuity in
$\Z_0$ of the functions $(W,W_\tau)$ defined by \equ(vs). In Proposition~2.2
below, we shall show that the Cauchy problem for \equ(uv) in $\Z_t$ is 
locally well-posed. 

Before stating our main result, we explain its content in a heuristic way. 
Taking formally the limit $t \to +\infty$ in \equ(uv) and using \equ(gamasym),
we see that $u$ satisfies the linear parabolic equation
$$
   u_t \,=\, \LL_\infty u \,\eqdef\, u_{xx} + \Bigl({x \over 2} +
   {2 \over x}\Bigr)u_x + {3 \over 2} u \quad \hbox{if } x > 0~, \quad
   u_x \,=\, 0 \quad \hbox{if } x \le 0~. \EQ(limiting)
$$
Therefore, it is reasonable to expect that the long-time behavior of
the solutions of \equ(uv) is determined by the spectral properties 
of the operator $\LL_\infty$ on $\real_+$, with Neumann boundary 
condition at $x=0$. Now, as is easily verified, this limiting operator
is just the image under the scaling \equ(vs) of the radially symmetric 
Laplace operator in three dimensions. Indeed, if $u$ and $W$ 
are related through \equ(vs), the equation $u_t = \LL_\infty u$ is 
equivalent to $W_\tau = W_{\xi\xi} + (2/\xi)W_{\xi}$, $\xi > 0$. 
This crucial observation explains the factor $(\tau{+}\tau_0)^{-3/2}$
in \equ(vs), and allows to compute exactly the spectrum of $\LL_\infty$ 
in various function spaces, see [\refnb{GR2}, Appendix~A]. For instance, 
in the space $\H^1(\real_+,(1{+}x)^6{\rm d}x)$, the spectrum of $\LL_\infty$
consists of a simple, isolated eigenvalue at $\lambda=0$, and of ``continuous''
spectrum filling the half-plane $\{\lambda \in \complex\,|\, \Re
\lambda \le -1/4\}$. The eigenfunction corresponding to $\lambda = 0$ 
is the Gaussian $\e^{-x^2/4}$. Therefore, we expect that the solution 
$u(x,t)$ of \equ(uv) converges as $t \to +\infty$ to $\alpha \phi^*(x)$ for 
some $\alpha \in \real$, where 
$$
   \phi^*(x) \,=\, {1 \over \sqrt{4\pi}} \cases{1 & if $x < 0$~,\cr
   \e^{-x^2/4} & if $x \ge 0$~.} \EQ(phistar)
$$
This function is normalized so that $\int_0^\infty x^2 \phi^*(x)\d x =
1$. Since $v = u_t - {x \over 2}u_x -{3 \over 2}u$, we also expect that 
$v(x,t)$ converges to $\alpha \psi^*(x)$, where $\psi^* = -{x \over 2}
\phi^*_x -{3 \over 2}\phi^*$. It is crucial to note that Eq.\equ(limiting)
is independent of $\epsilon$: this explains why the solutions of \equ(W), 
hence of \equ(U), behave for large times like those of the corresponding 
parabolic equations.

Our main result shows that these heuristic considerations are indeed 
correct:

\CLAIM Theorem(main) Assume that the nonlinearity $\FF$ satisfies 
\equ(nonlin), and let $\epsilon > 0$. There exist $t_0 > 0$, $\delta_0 > 0$ 
and $C > 0$ such that, for all initial data $(u_0,v_0) \in \Z_{t_0}$ with 
$\Phi_\eta(t_0,u_0,v_0) \le \delta_0^2$, the system \equ(uv) has a unique 
solution $(u,v) \in \CC([t_0,+\infty),\Z_t)$ satisfying $(u(t_0),v(t_0)) = 
(u_0,v_0)$. In addition, there exists $\alpha^* \in \real$ such that, for all
$t \ge t_0$,
$$\eqalign{
   \|u(t)-\alpha^*\phi^*\|^2_{\H^1_t} + \eta \e^{-t} \|v(t)-\alpha^*
    \psi^*\|^2_{\L^2_t} + \int_{t_0}^t \e^{-(t-s)/2}
    \|v(s)-\alpha^*\psi^*\|^2_{\L^2_s}\d s& \cr 
   \le\, C(1+t)^2\e^{-t/2}\Phi_\eta(t_0,u_0,v_0)&~.} \EQ(mainuv)
$$

\REMARKS \HB
{\bf 1.} In the proof of \clm(main), we shall take for convenience the 
parameter $t_0 = \log(\tau_0)$ large enough, but this choice is irrelevant 
since, as reflected in Corollary~1.3 below, the results for the original 
equation \equ(U) are not affected. 

\noindent{\bf 2.} The estimate \equ(mainuv) shows in particular that the 
solution $u(t)$ converges to $\alpha^* \phi^*$ like $t\e^{-t/4}$ as $t \to 
+\infty$. As was already mentioned, the decay rate $\e^{-t/4}$ corresponds to 
the spectral gap of the linear operator $\LL_\infty$ in $\H^1(\real_+,(1{+}x)^6
\d x)$, and is thus optimal in our function space. The same argument suggests
that this rate could be improved up to $\e^{-t/2}$ at the expense of assuming
a faster decay of $u,v$ as $x \to +\infty$, as in \ref{Ga}. 

\noindent{\bf 3.} \clm(main) does not give a satisfactory estimate
of the term $\|v(t)-\alpha^* \psi^*\|^2_{\L^2_t}$. If $\epsilon$ is
sufficiently small, using three additional pairs of functionals as
in Section~3, one can show that $\int_0^\infty (x{+}x^2) |v(x,t)-\alpha^* 
\psi^*(x)|^2\d x$ decays at least like $(1+t)^2\e^{-t/2}$ and that 
$\int_{-\infty}^0 \e^{2\kappa x\e^{t/2}} |v(x,t)-\alpha^* \psi^*(x)|^2 \d x +
\int_0^\infty |v(x,t)-\alpha^* \psi^*(x)|^2\d x$ is bounded by a polynomial
in $t$. Since these estimates are probably not optimal and were obtained for 
small $\epsilon$ only, the corresponding calculations will not be given here.

\noindent{\bf 4.} Given $\epsilon_0 > 0$ and a nonlinearity $\FF$ satisfying
\equ(nonlin), it is straightforward to verify that all the statements in the 
sequel (and their proofs) hold uniformly in $\epsilon$ for $\epsilon \in (0,
\epsilon_0]$. In particular, the constants $t_0, \delta_0, C$ appearing in 
\clm(main) are independent of $\epsilon$ for $\epsilon \in (0,\epsilon_0]$.  
As a consequence, taking the limit $\epsilon \to 0$ in \equ(mainuv), we obtain
a local stability result for the travelling waves of the parabolic equation 
\equ(U) with $\epsilon = 0$. Except for the use of slightly different 
function spaces, this result coincides with Theorem~1.1 of \ref{Ga}.

Combining \clm(main) and Lemma~2.4 below, we obtain in particular the
following convergence result for the perturbation in the original variables:

\CLAIM Corollary(W) Under the assumptions of \clm(main), the following 
estimate holds:
$$
   \sup_{\xi \in \real}(1+\e^{-\kappa\xi})^{-1} \left|W(\xi,\tau)
    -{\alpha^* \over \tau^{3/2}}\,\phi^*\Bigl({\xi \over \sqrt{\tau}}\Bigr)
    \right| \,=\, \OO(\tau^{-7/4}\log \tau)~,
$$
as $\tau \to +\infty$, where $W(\xi,\tau)$ is given by \equ(vs). 
Equivalently,
$$
   \sup_{\xi \in \real} \left(1 + {\e^{c_*\xi/2} \over 1+|\xi|}\right)
    \left|w(\xi,\t)-{\alpha \over \t^{3/2}}\,h'(\xi)\phi^*\Bigl({\xi
    \sqrt{1{+}\epsilon c_*^2} \over 
    \sqrt{\t}}\Bigr)\right| \,=\, \OO(\t^{-7/4}\log \t)~,
$$
as $\t \to +\infty$, where $\alpha = \alpha^* (1+\epsilon c_*^2)^{3/2}$
and $w(\xi,\t)$ is given by \equ(pert).

The rest of this paper is devoted to the proof of \clm(main), which is 
organized as follows. First of all, we prove that the Cauchy problem for
Eq.\equ(uv) is locally well-posed in the space $\Z_t$, in the sense of 
Definition~1.1. Then, in Section~2.1, we decompose the solutions $(u,v)$
of \equ(uv) using an approximate spectral projection of the time-dependent 
operator $\LL_t$ defined in (2.3) below. The first term in this 
decomposition is one-dimensional and converges to $\alpha^*(\phi^*,\psi^*)$
as $t \to +\infty$. The remainder $(f,g)$ satisfies an evolution system 
similar to \equ(uv), with additional terms which are estimated in Section~2.2. 
The core of the proof is Section~3, where the evolution of $(f,g)$ in $\Z_t$
is controlled using a hierarchy of energy functionals. As in \ref{GR2},
some of these quantities are constructed in terms of the primitives
$(F,G)$ rather than the functions $(f,g)$ themselves. Finally, the results 
are summarized in the short Section~4. 

Although the proof we present here is certainly not simple, we believe
that our approach is systematic and very well adapted to study the 
long-time asymptotics in a large class of dissipative systems. As a matter 
of fact, the present proof follows exactly the same lines as in \ref{GR2}, 
although the problems considered are significantly different. When compared 
with other accurate techniques, such as the Renormalization Group used in 
\ref{BK1} and \ref{Ga}, our method shows at least two advantages. First, 
we do not need precise estimates of the resolvent of the linearized operator 
around the travelling wave (although some spectral information is used to
construct our energy functionals). This substantial simplification is 
especially interesting in the perspective of possible applications to 
higher-dimensional problems, where standard tools like the Evans function 
are not available. Next, while most of our effort is devoted to controlling
the linear terms in \equ(uv), the nonlinearities are naturally incorporated 
into the scheme and do not require any extra argument. In the present case, 
the factor $\e^{-t/2}$ in front of the last term in \equ(uv) clearly shows 
that the nonlinearity is irrelevant for the long-time behavior, provided the 
solution $u(t)$ stays globally bounded. On the other hand, a minor drawback of
our approach is the introduction of non-autonomous systems and time-dependent 
function spaces through the change of variables \equ(sv). We shall avoid 
this difficulty by returning to the original variables to show that the 
Cauchy problem for \equ(uv) is locally well-posed and to prove that our 
energy functionals are differentiable in time. 

\LIKEREMARK(Notation) In the sequel, we denote by $C$ a generic positive
constant which may differ from place to place, while numbered 
constants $C_i, K_i, \dots$ keep the same value throughout the paper. 


\SECTION Preliminaries

We begin with a local existence result for the solutions $W$ of \equ(W) in 
the function space $\Z_0 = \H^1_0 \times \L^2_0$. We recall that $\H^1_0$, 
$\L^2_0$ are defined by the norms \equ(norms) with $t = 0$. 

\CLAIM Lemma(localW) Let $\eta > 0$ and $\delta > 0$. There exists 
$\ttau > 0$ such that, for all initial data $(W_0,\dot W_0) \in \Z_0$ 
with $\|(W_0,\dot W_0)\|_{\Z_0} \le \delta$, Eq.\equ(W) has a unique 
(mild) solution $W \in \CC([0,\ttau],\H^1_0) \cap \CC^1([0,\ttau],\L^2_0)$ 
satisfying $(W(0),W_\tau(0)) = (W_0,\dot W_0)$. The solution $(W,W_\tau)$ 
depends continuously on the initial data in $\Z_0$, uniformly in $\tau \in 
[0,\ttau]$. In addition, if $(W_0,\dot W_0) \in \H^2_0 \times \H^1_0$, then 
$W \in \CC([0,\ttau],\H^2_0) \cap \CC^1([0,\ttau],\H^1_0) \cap 
\CC^2([0,\ttau],\L^2_0)$ is a classical solution of Eq.\equ(W) in $\L^2_0$.

\PROOF Let $q \in \CC^\infty(\real)$ be a positive function 
satisfying $q(\xi) = \e^{-\kappa \xi}$ for $\xi \le 0$ and $q(\xi) = \xi^{-3}$
for $\xi \ge 1$. Setting $W(\xi,\tau) = q(\xi)\omega(\xi,\tau)$ in \equ(W), 
we obtain for $\omega$ the equation
$$
   \eta \omega_{\tau\tau} -2\nu \omega_{\xi\tau} \,=\, \omega_{\xi\xi}
   + \MM(\omega,\omega_\xi,\omega_\tau)~, \EQ(omega)
$$
where
$$\eqalign{
  \MM(\omega,\omega_\xi,\omega_\tau) \,&=\, -\left(1 - \nu\gamma 
   -2\nu{q' \over q}\right)\omega_\tau + \left(\gamma + {2q' \over q}\right)
   \omega_{\xi} + \left(\gamma {q' \over q} + {q'' \over q}\right)\omega \cr
  \,&+\, h' q \omega^2 \NN(h,h'q \omega)~.} \EQ(MM)
$$
Since the functions $\gamma$, $q'/q$, $q''/q$ and $h'q$ are all bounded, 
and since the nonlinearity $\FF$ in \equ(U) is $\CC^2$, it is straightforward
to verify that the map $\MM : \H^1(\real) \times \L^2(\real) \to \L^2(\real)$
defined by $(\omega,\omega_\tau) \mapsto \MM(\omega,\omega_\xi,\omega_\tau)$
is locally Lipschitz, uniformly on bounded subsets. 
Therefore, by a classical result \ref{CH}, the Cauchy problem for \equ(omega)
is locally well-posed in $\H^1 \times \L^2$. More precisely, for any $r > 0$, 
there exists $\ttau > 0$ such that, for all initial data $(\omega_0,
\dot \omega_0) \in \H^1 \times \L^2$ with $\|(\omega_0,\dot \omega_0)
\|_{\H^1 \times \L^2} \le r$, \equ(omega) has a unique (mild) solution 
$\omega \in \CC([0,\ttau],\H^1) \cap \CC^1([0,\ttau],\L^2)$ satisfying 
$(\omega(0),\omega_\tau(0)) = (\omega_0,\dot \omega_0)$. This solution 
depends continuously on the initial data in $\H^1 \times \L^2$, uniformly 
in $\tau \in [0,\ttau]$. 
Moreover, if $(\omega_0,\dot \omega_0) \in \H^2 \times \H^1$, then 
$\omega \in \CC([0,\ttau],\H^2) \cap \CC^1([0,\ttau],\H^1) \cap 
\CC^2([0,\ttau],\L^2)$ is a classical solution of Eq.\equ(omega). Thus, 
returning to the original function $W = q\omega$ and using the fact that
$$
   C^{-1} \|(\omega,\omega_\tau)\|_{\H^1 \times \L^2} \,\le\, 
   \|(W,W_\tau)\|_{\Z_0} \,\le\, C\|(\omega,\omega_\tau)\|_{\H^1 \times \L^2}~,
$$
for some $C \ge 1$, we obtain the desired result, if $r = C\delta$. This 
concludes the proof of \clm(localW). \QED

As a consequence of Definition~1.1 and \clm(localW), we obtain the following
existence result for the solution $(u,v)$ of \equ(uv):

\CLAIM Proposition(localuv) Let $\eta > 0$, $\delta_1 > 0$, $t_2 > t_0$. 
There exists $T > 0$ such that, for all $t_1 \in [t_0,t_2]$ and all 
$(u_1,v_1) \in \Z_{t_1}$ satisfying $\Phi_\eta(t_1,u_1,v_1) \le \delta_1^2$, 
the system \equ(uv) has a unique solution $(u,v) \in \CC([t_1,t_1+T],\Z_t)$ 
with initial data $(u(t_1),v(t_1)) = (u_1,v_1)$. 

\REMARK In particular, \clm(localuv) implies that, if $(u,v) \in \CC([t_0,
t_*),\Z_t)$ is a maximal solution of \equ(uv) and if $\Phi_\eta(t,u(t),v(t))
\le \delta_1^2$ for all $t \in [t_0,t_*)$, then actually $t_* = +\infty$, 
\ie the solution $(u,v)$ is globally defined. 

\PROOF Given $t_1 \in [t_0,t_2]$ and $(u_1,v_1) \in \Z_{t_1}$ satisfying
$ \Phi_\eta(t_1,u_1,v_1) \le \delta_1^2$, we define
$$
   W_1(\xi) \,=\, \e^{-3t_1/2} u_1(\xi \e^{-t_1/2})~, \quad
   \dot W_1(\xi) \,=\, \e^{-5t_1/2} v_1(\xi \e^{-t_1/2})~,
   \quad \xi \in \real~.
$$
Then $(W_1,\dot W_1) \in \Z_0$, and there exists a constant $C > 0$ 
(depending on $\eta$ and $t_2$) such that $\|(W_1,\dot W_1)\|_{\Z_0} \le 
C\delta_1$. Since Eq.\equ(W) is autonomous, it follows from \clm(localW) 
that there exists a time $\ttau > 0$, depending on $\eta$, $C\delta_1$ but not
on $(W_1,\dot W_1)$, such that \equ(W) has a unique (mild) solution 
$W \in \CC([\e^{t_1},\e^{t_1}{+}\ttau],\H^1_0) \cap \CC^1([\e^{t_1},\e^{t_1}{+}
\ttau],\L^2_0)$ satisfying $W(\xi,\e^{t_1}) = W_1(\xi)$, $W_\tau(\xi,\e^{t_1}) 
= \dot W_1(\xi)$. Now, we set $T = \log(1+\ttau \e^{-t_2})$, and for all 
$t \in [t_1,t_1+T] \subset [t_1,\log(\e^{t_1}+\ttau)]$ we define
$$
   u(x,t) \,=\, \e^{3t/2} W(x\e^{t/2},\e^t)~, \quad
   v(x,t) \,=\, \e^{5t/2} W_\tau(x\e^{t/2},\e^t)~.
$$
By Definition~1.1, $(u,v) \in \CC([t_1,t_1+T],\Z_1)$ is a solution of \equ(uv) 
with $(u(t_1),v(t_1)) = (u_1,v_1)$, and the uniqueness of this 
solution follows from the uniqueness of $W$ as a mild solution of \equ(W).
This concludes the proof of \clm(localuv). \QED

\SUBSECTION Spectral Decomposition of the Solution

{}From now on, we assume that $(u,v) \in \CC([t_0,t_1],\Z_t)$ is a solution of 
\equ(uv) in the sense of \clm(localuv). Inspired by \ref{Ga} and \ref{GR2},
we shall decompose this solution using an approximate spectral projection
of the (time-dependent) linear operator
$$
   \LL_t \,=\, \partial_x^2 + \left({x \over 2} + \e^{t/2}
   \gamma(x\e^{t/2})\right)\partial_x + {3 \over 2}~, \EQ(opLLt)
$$
which appears in \equ(uv). As is easily verified, the function $\phi^*$ 
defined in \equ(phistar) is an approximate eigenfunction of $\LL_t$, in the 
sense that $\|\LL_t \phi^*\|_{\L^2_t} = \OO(\e^{-t/4})$ as $t \to +\infty$. 
The corresponding approximate spectral projection in $\L^2_t$ is given by 
the formula
$$
   u \mapsto \left(\int_\real \e^{-t}p(x\e^{t/2})u(x)\d x\right)\phi^*
   \EQ(projection)
$$
where $p:\real \to \real$ is the (unique) solution of the differential 
problem
$$
   p'(\xi) \,=\, \gamma(\xi)p(\xi)~, \quad \xi \in \real~,
   \quad \lim_{\xi \to +\infty} {p(\xi) \over \xi^2} \,=\, 1~. \EQ(pdiff)
$$
It follows from \equ(gamasym), \equ(pdiff) that $p(\xi) > 0$ for all $\xi 
\in \real$, and $p(\xi) = \OO(\e^{\gamma_-\xi})$ as $\xi \to -\infty$.

Motivated by \equ(projection), we introduce the functions
$$
   \phi(x,t) \,=\, {\phi^*(x) \over 1+\zeta(t)}~, \quad 
   \psi(x,t) \,=\, \phi_t(x,t) - {x \over 2}\phi_x(x,t) - {3 \over 2}
   \phi(x,t)~, \EQ(phipsi)
$$
where
$$
   \zeta(t) \,=\, \int_\real \e^{-t} p(x\e^{t/2}) \phi^*(x)\d x - 1~.
   \EQ(zeta)
$$
We shall show in the proof of Lemma~2.5 below that $\zeta(t)$ and $\zeta'(t)$ 
converge to zero as $t \to +\infty$, so that $\phi(x,t) \to \phi^*(x)$ and 
$\psi(x,t) \to \psi^*(x)$. By construction, we also have
$$
   \int_\real \e^{-t} p(x\e^{t/2}) \phi(x,t)\d x \,=\, 1~, \quad
   \int_\real \e^{-t} p(x\e^{t/2}) \psi(x,t)\d x \,=\, 0~, \quad t \ge 0~.
   \EQ(orth)
$$
Using these notations, we decompose the solution $(u,v)$ of \equ(uv) as
$$
   u(x,t) \,=\, \alpha(t)\phi(x,t) + f(x,t)~, \quad
   v(x,t) \,=\, \beta(t)\phi(x,t) + \alpha(t) \psi(x,t) + g(x,t)~,
   \EQ(decomp)
$$
where 
$$
   \alpha(t) \,=\, \int_\real \e^{-t}p(x\e^{t/2}) u(x,t)\d x~, \quad
   \beta(t) \,=\,  \int_\real \e^{-t}p(x\e^{t/2}) v(x,t)\d x~. \EQ(alpha)
$$
In view of \equ(orth), \equ(alpha), the functions $f,g$ satisfy the 
``orthogonality relations''
$$
   \int_\real \e^{-t} p(x\e^{t/2}) f(x,t)\d x \,=\, 0~, \quad
   \int_\real \e^{-t} p(x\e^{t/2}) g(x,t)\d x \,=\, 0~.
   \EQ(intfg)
$$
We now determine the evolution equations satisfied by $\alpha,\beta,f,g$. 
Our first result is:

\CLAIM Lemma(alpha) If $(u,v) \in \CC([t_0,t_1],\Z_t)$ is a solution
of \equ(uv), then $\alpha \in \CC^2([t_0,t_1])$ and
$$
  {\dd \over \dd t}\alpha(t) \,=\, \beta(t)~, \quad  
  {\dd \over \dd t}\left(\eta \e^{-t}\beta(t)+\alpha(t)\right) 
  \,=\, m(t)~, \EQ(dotalpha)
$$
where
$$
   m(t) \,=\, \int_\real \e^{-t} p(x\e^{t/2}) \left(-\nu \gamma(x\e^{t/2})
   v(x,t) + \e^{-t/2} h'(x\e^{t/2}) u(x,t)^2 N(x,t)\right) \d x~.
$$

\PROOF Let $\tau_1 = \e^{t_1}-\tau_0$, and let $W(\xi,\tau)$ be given
by \equ(vs) for $\tau \in [0,\tau_1]$. By Definition~1.1, $W \in 
\CC([0,\tau_1],\H^1_0) \cap \CC^1([0,\tau_1],\L^2_0)$ is a (mild) 
solution of \equ(W). Since $\alpha(t) = \int_\real p(\xi)W(\xi,\e^t-\tau_0)
\d \xi$, it follows that $\alpha \in \CC^1([t_0,t_1])$ and
$$
   {\dd \over \dd t}\alpha(t) \,=\, 
   \e^t \int_\real p(\xi)W_\tau(\xi,\e^t-\tau_0)\d \xi \,=\, 
   \int_\real \e^{-t} p(x\e^{t/2})v(x,t)\d x \,=\, \beta(t)~.
$$
To prove that $\alpha \in \CC^2([t_0,t_1])$, we first assume that 
$(W_0,\dot W_0) \equiv (W(\cdot,0),W_\tau(\cdot,0)) \in \H^2_0 \times \H^1_0$.
Then, by \clm(localW), $W \in \CC([0,\tau_1],\H^2_0) \cap \CC^1([0,\tau_1],
\H^1_0) \cap \CC^2([0,\tau_1],\L^2_0)$ is a classical solution of \equ(W), 
hence $\alpha \in \CC^2([t_0,t_1])$ and
$$
   {\dd \over \dd t}(\eta\e^{-t}\beta(t)+\alpha(t)) \,=\, \e^t \int_\real 
   p(\xi) (\eta W_{\tau\tau} + W_\tau)(\xi,\e^t-\tau_0)\d \xi
   \,\eqdef\, m(t)~.
$$
Since $p(\eta W_{\tau\tau}+W_\tau) = (pW_\xi)_\xi + 2\nu (pW_\tau)_\xi
-\nu p \gamma W_\tau + ph' W^2\NN(h,h'W)$ by \equ(W), \equ(pdiff), we find
$$\eqalign{
  m(t) \,&=\, \e^t \int_\real p(\xi)\left(-\nu \gamma W_\tau + 
   h' W^2\NN(h,h'W)\right)(\xi,\e^t-\tau_0)\d \xi \cr
  \,&=\, \int_\real \e^{-t}p(x\e^{t/2})\left(-\nu \gamma(x\e^{t/2}) 
   v(x,t) + \e^{-t/2}h'(x\e^{t/2}) u(x,t)^2 N(x,t)\right)\d x~.}
$$
For all $t \in [t_0,t_1]$, we thus have
$$
  \eta \e^{-t}\beta(t)+\alpha(t) \,=\, \eta \e^{-t_0}
  \beta(t_0)+\alpha(t_0) + \int_{t_0}^t m(s)\d s~. \EQ(integ)
$$
By \clm(localW), both sides of \equ(integ) are continuous functions of the 
initial data $(W_0,\dot W_0)$ in $\Z_0$. Since \equ(integ) is satisfied for 
all $(W_0,\dot W_0)$ in the dense subspace $\H^2_0 \times \H^1_0$, the 
equality must hold for all $(W_0,\dot W_0) \in \Z_0$. This shows that $\eta 
\e^{-t} \beta+\alpha \in \CC^1([t_0,t_1])$ and that \equ(dotalpha) holds. 
The proof of \clm(alpha) is complete. \QED

It follows from \equ(uv), \equ(decomp) and \clm(alpha) that $(f,g) \in 
\CC([t_0,t_1],\Z_t)$ is a solution (in the sense of Definition~1.1) of the 
system
$$ \eqalign{
   &f_t - {x \over 2} f_x - {3 \over 2} f \,=\, g~, \cr
   &\eta \e^{-t}\bigl(g_t - {x \over 2}g_x - {5 \over 2}g\bigr) + 
   (1-\nu\gamma(x \e^{t/2}))g -2\nu\e^{-t/2}g_x \,=\, \cr
   &\qquad f_{xx} + \e^{t/2} \gamma(x\e^{t/2}) f_x + r(x,t)~,} \EQ(fg)
$$
where
$$\eqalign{
  r(x,t) \,=\, &\alpha(\phi_{xx}+\e^{t/2}\gamma(x\e^{t/2})\phi_x -\psi)
   -\eta \e^{-t}(2\beta\psi + \alpha(\psi_t-\frac{x}{2}\psi_x -
   \frac{5}{2}\psi)) \cr
   &+ \nu\gamma(x\e^{t/2})(\beta\phi+\alpha\psi) + 2\nu\e^{-t/2}
    (\beta\phi_x+\alpha\psi_x)\cr 
   &+ \e^{-t/2} h'(x\e^{t/2}) u(x,t)^2 N(x,t) - m(t)\phi~.} \EQ(remainder)
$$
Using \equ(pdiff), \equ(phipsi), \equ(orth) and the definition of $m(t)$ in 
\clm(alpha), it is not difficult to verify that
$$
  \int_\real \e^{-t} p(x\e^{t/2}) \bigl(r(x,t) - \nu\gamma(x\e^{t/2})g(x,t)
  \bigr) \d x \,=\, 0~. \EQ(intr)
$$

Finally, as in \ref{GR2}, it will be useful to consider also the primitives 
$$
  F(x,t) \,=\, \int_{-\infty}^x \e^{-t} p(y\e^{t/2}) f(y,t) \d y~, \quad
  G(x,t) \,=\, \int_{-\infty}^x \e^{-t} p(y\e^{t/2}) g(y,t) \d y~.
  \EQ(primitives)
$$
Using \equ(intfg) and standard inequalities (see Lemma~2.7 below and the 
remark at the end of this section), it is straightforward to verify that 
$(F,G) \in \CC^1([t_0,t_1],\H^1\times\L^2)$ is a classical solution of the 
system
$$\eqalign{
   &F_t - {x \over 2} F_x \,=\, G~, \cr
   &\eta \e^{-t}\bigl(G_t - {x \over 2}G_x - G\bigr) + 
   G -2\nu\e^{-t/2}G_x \,=\, F_{xx} - \e^{t/2} \gamma(x\e^{t/2}) F_x 
   + R(x,t)~,} \EQ(FG)
$$
where
$$
  R(x,t) \,=\, \int_{-\infty}^x \e^{-t} p(y\e^{t/2}) \bigl(r(y,t) 
  -\nu\gamma(y\e^{t/2})g(y,t)\bigr) \d y~. \EQ(primR)
$$

\SUBSECTION Bounds on the Nonlinear Terms

In this subsection, we assume that $(u,v) \in \CC([t_0,t_1],\Z_t)$ is a 
solution of \equ(uv) satisfying the bound
$$
   \|u(t)\|_{\H^1_t} \,\le\, 1~, \quad t \in [t_0,t_1]~. \EQ(ubound)
$$
Then $u(t)$ is uniformly bounded in a weighted $\L^\infty$ space, as
a consequence of the following result:

\CLAIM Lemma(Linfty) There exists a constant $K_0 > 0$ such that, 
for all $t \ge 0$ and all $w \in \H^1_t$, 
$$
   \sup_{x \le 0} \e^{\kappa x\e^{t/2}}|w(x)| + \sup_{x\ge 0}
   (1{+}x)^3|w(x)| \,\le\, K_0 \|w\|_{\H^1_t}~. \EQ(embedding)
$$

\REMARK Note the crucial fact that the constant $K_0$ in \equ(embedding)
is independent of $t$. 
 
\PROOF Let $t \ge 0$ and $w \in \H^1_t$. By a classical inequality, 
there exists $C > 0$ such that
$$ \sup_{x\ge 0}(1{+}x)^6|w(x)|^2 \,\le\, C \int_0^\infty (1{+}x)^6 
   (w(x)^2 +w'(x)^2)\d x~. \EQ(wfirst)
$$
In particular, $w(0)^2 \le C\|w\|_{\H^1_t}^2$. On the other hand, we have
for all $x < 0$:
$$\eqalign{
  \e^{2\kappa x\e^{t/2}}w(x)^2 \,&=\, w(0)^2 -\int_x^0 \e^{2\kappa y\e^{t/2}}
   \bigl(2w(y)w'(y) +2\kappa \e^{t/2}w(y)^2\bigr)\d y \cr
  \,&\le\, w(0)^2 + \int_{-\infty}^0 \e^{2\kappa y\e^{t/2}} (w(y)^2 + 
   w'(y)^2)\d y~.} \EQ(wsecond)
$$
Combining \equ(wfirst), \equ(wsecond), we obtain \equ(embedding). This 
concludes the proof of \clm(Linfty). \QED

In the sequel, it will be natural to control the solution $(u,v)$ of
\equ(uv) in terms of the functions $\alpha, \beta, f, g$ defined in 
\equ(decomp), \equ(alpha). The equivalence of the corresponding norms is the 
content of our next result:

\CLAIM Lemma(equiv) There exists a constant $K_1 \ge 1$ such that, for 
all $t \ge 0$ and all $(u,v) \in \Z_t$, 
$$\eqalign{
  K_1^{-1} \|u\|_{\H^1_t} \,&\le\, |\alpha| + \|f\|_{\H^1_t} 
   \,\le\, K_1 \|u\|_{\H^1_t}~, \cr
  K_1^{-1} \|v\|_{\L^2_t} \,&\le\, |\alpha| + |\beta|
   + \|g\|_{\L^2_t} \,\le\, K_1
   (\|u\|_{\H^1_t} +\|v\|_{\L^2_t})~,} \EQ(K1bound)
$$
where $\alpha, \beta$ are defined in \equ(alpha) and $f,g$ in \equ(decomp). 

\PROOF From \equ(gamasym), we know that $\gamma(\xi) \to \gamma_-$ as 
$\xi \to -\infty$ and $\gamma(\xi) \sim 2/(\xi{+}\xi_0)$ as $\xi \to 
+\infty$. Setting $\xi_1 = -\xi_0 +2/\gamma_-$, we decompose $\gamma(\xi)$ 
as $\gamma_0(\xi) + \hat \gamma(\xi)$, where
$$
   \gamma_0(\xi) \,=\, \cases{ \gamma_- & if $\xi < \xi_1$~, \cr
   2/(\xi{+}\xi_0) & if $\xi \ge \xi_1$~.}
$$
By \equ(gamasym), the remainder $\hat \gamma(\xi)$ decays exponentially
as $|\xi| \to \infty$. Thus the solution of \equ(pdiff) can be represented
as
$$
   p(\xi) = p_0(\xi) \exp\left(-\int_\xi^\infty \hat \gamma(s)\d s\right)
   \,, \quad  p_0(\xi) = \cases{ (2/\gamma_-)^2 \e^{\gamma_- (\xi-\xi_1)} 
   & if $\xi < \xi_1$, \cr (\xi{+}\xi_0)^2 & if $\xi \ge \xi_1$.} \EQ(prep)
$$
In particular, there exists $C_0 \ge 1$ such that
$$
   p(\xi) \,\le\, C_0 \cases{\e^{\gamma_- \xi} & if $\xi < 0$~, \cr
   (1{+}\xi)^2 & if $\xi \ge 0$~,} \quad 
   p(\xi) \,\ge\, C_0^{-1} \cases{\e^{\gamma_- \xi} & if $\xi < 0$~, \cr
   (1{+}\xi)^2 & if $\xi \ge 0$~.} \EQ(pbound)
$$

Using \equ(prep) and remembering that $\int_0^\infty x^2 \phi^*(x) \d x = 1$, 
we decompose the function $\zeta(t)$ defined in \equ(zeta) as
$$\eqalign{
  \zeta(t) \,&=\, \int_{-\infty}^0 \e^{-t}p(x\e^{t/2})\phi^*(x)\d x + 
   \int_0^\infty \e^{-t}\bigl(p(x\e^{t/2})-p_0(x\e^{t/2})\bigr)\phi^*(x)\d x 
   \cr
  \,&+\, \int_0^\infty \bigl(\e^{-t}p_0(x\e^{t/2})-x^2\bigr)\phi^*(x)\d x
   \,=\, \zeta_1(t) + \zeta_2(t) + \zeta_3(t)~.}
$$
Using \equ(phistar), we remark that
$$
  \zeta_1(t) = {\e^{-3t/2} \over \sqrt{4\pi}} \int_{-\infty}^0 p(\xi)\d \xi~,
  \quad  \zeta_2(t) \,=\, \e^{-3t/2} \int_0^\infty \bigl(p(\xi)-p_0(\xi)
  \bigr) \phi^*(\xi \e^{-t/2})\d \xi~,
$$
where $p(\xi)-p_0(\xi)$ decays exponentially to zero as $\xi \to +\infty$ due 
to \equ(prep). On the other hand, setting $\bar \xi = \max(0,\xi_1)$, we have
$$
   \zeta_3(t) \,=\, \e^{-3t/2} \int_0^{\bar \xi} \bigl(p_0(\xi)-
   (\xi{+}\xi_0)^2\bigr)\phi^*(\xi \e^{-t/2})\d \xi + 
   \int_0^\infty \bigl(2\xi_0 x \e^{-t/2} + \xi_0^2 \e^{-t}\bigr)
   \phi^*(x)\d x~.
$$
It follows immediately from these expressions that 
$$
   |\zeta(t)| + |\zeta'(t)| + |\zeta''(t)| \,\le\, C_1 \e^{-t/2}~,
   \quad (1+\zeta(t))^{-1} \,\le\, C_1~, \quad t \ge 0~, \EQ(zetadot)
$$
for some $C_1 > 0$. As a consequence, the functions $\phi(x,t), \psi(x,t)$ 
defined by \equ(phipsi) satisfy the bounds
$$
   \|\phi(t)\|_{\H^1_t} + \|\psi(t)\|_{\L^2_t} \,\le\, C_2~, \quad
   t \ge 0~, \EQ(phipsibd)
$$
and
$$
   \|\phi(t)-\phi^*\|_{\H^1_t} + \|\psi(t)-\psi^*\|_{\L^2_t} \,\le\, C_2
   \e^{-t/2}~, \quad t \ge 0~, \EQ(phipsilim)
$$
for some $C_2 > 0$. 

Now, let $t \ge 0$, $(u,v) \in \Z_t$, and let $\alpha,\beta$ be defined as
in \equ(alpha). In view of \equ(pbound), we have
$$\eqalign{
  |\alpha| \,&\le\, C_0 \int_0^\infty (1{+}x)^2|u(x)|\d x + 
   C_0 \e^{-t} \int_{-\infty}^0 \e^{\gamma_- x\e^{t/2}} |u(x)|\d x \cr
  \,&\le\, C_0\left(\int_0^\infty (1{+}x)^6 |u|^2\d x\right)^{1/2} + 
   {C_0 \e^{-5t/4} \over (\gamma_- {-} \kappa)^{1/2}} \left(\int_{-\infty}^0
   \e^{2\kappa x\e^{t/2}}|u|^2\d x\right)^{1/2}~,} \EQ(intbound1) 
$$
hence $|\alpha| \le C_3 \|u\|_{\L^2_t}$ for some $C_3 > 0$. Similarly, we 
have $|\beta| \le C_3 \|v\|_{\L^2_t}$. Using these bounds together with 
\equ(decomp), \equ(phipsibd), we obtain \equ(K1bound). This concludes the 
proof of \clm(equiv). \QED

We now estimate the remainder terms $m(t)$ and $r(x,t)$ in \equ(dotalpha), 
\equ(fg). 

\CLAIM Lemma(rmbound) There exists a constant $K_2 > 0$ such that, if
$(u,v) \in \CC([t_0,t_1],\Z_t)$ is a solution of \equ(uv) satisfying 
\equ(ubound), then
$$
   \|r(t)\|_{\L^2_t} + \e^{t/4}|m(t)| \,\le\, K_2 \e^{-t/4}
   \left(\alpha(t)^2 + \beta(t)^2 + \|f(t)\|_{\H^1_t}^2 + 
   \|g(t)\|_{\L^2_t}^2 \right)^{1/2}~,
   \EQ(rmbound)
$$
for all $t \in [t_0,t_1]$, where $r(x,t)$ is defined in \equ(remainder)
and $m(t)$ in \clm(alpha). 

\PROOF We first consider the function $r_1(x,t) = \phi_{xx} + \e^{t/2}
\gamma(x\e^{t/2})\phi_x - \psi$. It follows from \equ(phistar), \equ(phipsi) 
that $r_1(x,t) = (1+\zeta(t))^{-1} (\hat r(x,t) + \zeta'(t)\phi(x,t))$, 
where 
$$
   \hat r(x,t) \,=\, \cases{(\e^{t/2}\gamma(x\e^{t/2}) - 2/x)\phi^*_x & 
   if $ x > 0$~, \cr {3\phi^*/2} & if $x < 0$~.}
$$
By \equ(zetadot), \equ(phipsibd), we have $\|\zeta'(t)\phi(t)\|_{\L^2_t} 
\le C_1 C_2 \e^{-t/2}$. To bound $\hat r(x,t)$, we observe that the function 
$\xi \mapsto (2 - \xi\gamma(\xi))$ belongs to $\L^2(\real_+)$ by 
\equ(gamasym). Since $\phi^*_x = -(x/2)\phi^*$ for $x > 0$, we thus find
$$\eqalign{
  &\int_0^\infty (1{+}x)^6 \hat r(x,t)^2 \d x \,\le\, {\e^{-t/2} \over 4} 
   \left(\sup_{x \ge 0} (1{+}x)^6 \phi^*(x)^2\right) \int_0^\infty 
   (2-\xi\gamma(\xi))^2\d \xi~, \cr
  &\int_{-\infty}^0 \e^{2\kappa x\e^{t/2}} \hat r(x,t)^2\d x = {9 \over 
   32\pi\kappa} \,\e^{-t/2}~.}
$$
Summarizing, we obtain $\|r_1(t)\|_{\L^2_t} \le C_4 \e^{-t/4}$ for some 
$C_4 > 0$. Similarly, since $\gamma \in \L^2(\real_+) \cap \L^\infty
(\real_-)$, we find $\|\gamma(x\e^{t/2})\phi(t)\|_{\L^2_t} \le C_4 \e^{-t/4}$ 
and $\|\gamma(x\e^{t/2})\psi(t)\|_{\L^2_t} \le C_4 \e^{-t/4}$. 

We next bound the non-linear term $r_2(x,t) = \e^{-t/2} h'(x\e^{t/2})u(x,t)^2 
N(x,t)$, where $N(x,t) = \NN(h(x\e^{t/2}),\e^{-3t/2}h'(x\e^{t/2})u(x,t))$. 
In view of \equ(hasym), \equ(ubound), \equ(embedding), there exists 
$C_5 > 0$ such that $\sup_{x\in \real} |h'(x\e^{t/2})u(x,t)| \le C_5$ for 
all $t \in [t_0,t_1]$. In particular, since $\NN : \real^2 \to \real$ 
is continuous, we have $\|N(\cdot,t)\|_{\L^\infty} \le N_0$ for some 
$N_0 > 0$ and all $t \in [t_0,t_1]$. It follows that $\|r_2\|_{\L^2_t}
\le \e^{-t/2} C_5 N_0 \|u(t)\|_{\L^2_t}$ for $t \in [t_0,t_1]$. 

Finally, the function $m(t)$ defined in \clm(alpha) can be written as
$m_1(t) + m_2(t)$, where
$$
   m_1(t) \,=\, -\nu \int_\real \e^{-t} p(x\e^{t/2}) \gamma(x\e^{t/2})
   v(x,t) \d x~, \quad 
   m_2(t) \,=\, \int_\real \e^{-t} p(x\e^{t/2}) r_2(x,t)\d x~.
$$
Proceeding as in \equ(intbound1), we find $|m_2(t)| \le C_3 
\|r_2(t)\|_{\L^2_t} \le \e^{-t/2} C_3 C_5 N_0 \|u(t)\|_{\L^2_t}$. Moreover, 
since $\e^{-t}\gamma(x\e^{t/2})p(x\e^{t/2}) \le C\e^{-t/2}(1+x)$ for 
$x \ge 0$, we obtain
$$
  |m_1(t)| \,\le\, C\nu \e^{-t/2} \int_0^\infty (1{+}x)|v(x,t)|\d x +
   C \nu \e^{-t} \int_{-\infty}^0 \e^{\gamma_- x\e^{t/2}}|v(x,t)|\d x~,
$$
hence $|m_1(t)| \le C_6 \nu \e^{-t/2} \|v(t)\|_{\L^2_t}$ for some
$C_6 > 0$. Therefore, there exists $C_7 > 0$ such that
$$
   |m(t)| \le C_7 \e^{-t/2}(\|u(t)\|_{\L^2_t} + \|v(t)\|_{\L^2_t})~, 
     \quad t \in [t_0,t_1]~. \EQ(mbound)
$$

Summarizing our results and observing that the functions $\phi$, $\phi_x$, 
$\psi$, $\psi_x$, $\psi_t$, $x\psi_x$ are uniformly bounded in $\L^2_t$ by
\equ(phipsi), \equ(zetadot), we see that the remainder $r(x,t)$ defined by 
\equ(remainder) satisfies
$$
   \|r(t)\|_{\L^2_t} \,\le\, C_8 \e^{-t/4} \left(|\alpha(t)| + 
   |\beta(t)| + \|u(t)\|_{\H^1_t} + \|v(t)\|_{\L^2_t}\right)~,
   \quad t \in [t_0,t_1]~, \EQ(rbound)
$$
for some $C_8 > 0$. Combining \equ(K1bound), \equ(mbound), \equ(rbound),
we obtain \equ(rmbound). This concludes the proof of \clm(rmbound). \QED

Finally, we bound the primitives $F,G,R$ defined in \equ(primitives), 
\equ(primR).

\CLAIM Lemma(Fbound) There exists a constant $K_3 > 0$ such that, for
all $t \ge 0$ and all $f \in \L^2_t$ satisfying $\int_\real p(x\e^{t/2})
f(x)\d x = 0$, the following estimate holds
$$
  \int_\real \left(1+{\e^t \over p(x\e^{t/2})}\right) F^2 \d x 
   \,\le\, K_3 \left(\e^{-2t} \int_{-\infty}^0 \!\e^{2\kappa x\e^{t/2}}
   f^2\d x + \int_0^\infty (1{+}x)^6 f^2\d x\right), \EQ(Fbound)
$$
where $F(x) = \int_{-\infty}^x \e^{-t}p(y\e^{t/2})f(y)\d y$. 

\PROOF Let $t \ge 0$ and $f \in \L^2_t$. We start from the identity
$$
   \e^t \int_{-\infty}^0 \e^{-\gamma_- x\e^{t/2}} F(x)^2\d x + 
   {\e^{t/2} \over \gamma_-} F(0)^2 \,=\, {2\e^{-t/2} \over \gamma_-}
   \int_{-\infty}^0 \e^{-\gamma_- x\e^{t/2}} p(x\e^{t/2}) F(x)f(x)\d x~,
$$
which is a simple integration by parts. Applying H\"older's inequality to
the right-hand side, we obtain
$$
   \e^t \int_{-\infty}^0 \e^{-\gamma_- x\e^{t/2}} F(x)^2\d x + {\e^{t/2} 
   \over \gamma_-} F(0)^2 \,\le\, {4\e^{-2t} \over \gamma_-^2} 
   \int_{-\infty}^0 \e^{-\gamma_- x\e^{t/2}} p(x\e^{t/2})^2 f(x)^2\d x~.
$$
Using \equ(pbound) and remembering that $\gamma_- = c+2\kappa > 2\kappa$, we 
conclude that
$$
  \int_{-\infty}^0 \left(1 + {\e^t \over p(x\e^{t/2})}\right) F(x)^2\d x 
  + \e^{t/2} F(0)^2 \,\le\, C \e^{-2t} \int_{-\infty}^0 \e^{2\kappa 
  x\e^{t/2}}f(x)^2\d x~, \EQ(F1)
$$
for some $C > 0$. 

Since $\int_\real p(x\e^{t/2})f(x)\d x = 0$, we have $F(x) = -\int_x^\infty 
\e^{-t}p(y\e^{t/2})f(y)\d y$. Using \equ(pbound) and a classical inequality 
of Hardy [\refnb{HLP}, Theorem~328], we find
$$
   \int_0^\infty F(x)^2\d x \,\le\, 4 \int_0^\infty \e^{-2t}x^2 
   p(x\e^{t/2})^2f(x)^2 \d x \,\le\, 4 C_0^2 \int_0^\infty 
   (1{+}x)^6 f(x)^2 \d x~. \EQ(F2)
$$
On the other hand, since $F(x) = F(0) + \int_0^x \e^{-t}p(y\e^{t/2})
f(y)\d y$, we have for $x > 0$
$$
   {\e^{t/2} |F(x)| \over 1+x\e^{t/2}} \,\le\, {\e^{t/2} |F(0)| 
   \over 1+x\e^{t/2}} + {C_0 \over x}\int_0^x (1{+}y)^2 |f(y)|\d y~. \EQ(F3)
$$
Using another form of Hardy's inequality [\refnb{HLP}, Theorem~327], we 
thus obtain
$$
   \int_0^\infty {\e^t F(x)^2 \over (1{+}x\e^{t/2})^2}\d x
   \,\le\, 2\e^{t/2}|F(0)|^2 + 8C_0^2 \int_0^\infty (1{+}x)^4 
   f(x)^2\d x~. \EQ(F4)
$$
Combining \equ(F1), \equ(F2), \equ(F4) and using \equ(pbound), we 
arrive at \equ(Fbound). This concludes the proof of \clm(Fbound). \QED

\CLAIM Lemma(Rbound) There exists a constant $K_4 > 0$ such that, 
if $(u,v) \in \CC([t_0,t_1],\Z_t)$ is a solution of \equ(uv) satisfying
\equ(ubound), then
$$
  \int_\real \left(1+{\e^t \over p(x\e^{t/2})}\right) R^2 \d x 
   \,\le\, K_4 \e^{-t/2} \left(\alpha(t)^2 + \beta(t)^2 + \|f(t)\|_{\H^1_t}^2 
   + \|g(t)\|_{\L^2_t}^2\right)~, \EQ(Rbound)
$$
for all $t \in [t_0,t_1]$, where $R(x,t)$ is defined in \equ(primR).

\PROOF Following the proof of \clm(Fbound), we obtain as in \equ(F1)
$$
  \int_{-\infty}^0 \left(1 + {\e^t \over p(x\e^{t/2})}\right) R(x,t)^2\d x 
  + \e^{t/2} R(0,t)^2 \,\le\, C \e^{-2t} \int_{-\infty}^0 \e^{2\kappa 
  x\e^{t/2}}(r^2 + \nu^2 \gamma_-^2 g^2) \d x~. 
$$
Next, remarking that $\e^{-t}p(x\e^{t/2}) \gamma(x\e^{t/2}) \le 
C\e^{-t/2}(1{+}x)$ for $x \ge 0$, we find instead of \equ(F2), \equ(F4)
$$\eqalign{
  \int_0^\infty R(x,t)^2 \d x \,&\le\, C\int_0^\infty \left((1{+}x)^6 
   r(x,t)^2 + \nu^2 \e^{-t} (1{+}x)^4 g(x,t)^2\right)\d x~,\cr
  \int_0^\infty {R(x,t)^2 \over (1{+}x\e^{t/2})^2} \d x \,&\le\, 
\  2\e^{t/2}R(0,t)^2 + C\int_0^\infty \left((1{+}x)^4 r^2 + \nu^2 \e^{-t}
   (1{+}x)^2 g^2\right)\d x~.}
$$
Combining these estimates, we obtain
$$
   \int_\real \left(1+{\e^t \over p(x\e^{t/2})}\right) R(x,t)^2 \d x 
   \,\le\, C \left(\|r(t)\|_{\L^2_t}^2 + \nu^2 \e^{-t} \|g(t)\|_{\L^2_t}^2
   \right)~,
$$
and \equ(Rbound) follows using \clm(rmbound). This concludes the proof of 
\clm(Rbound). \QED

\REMARK For $t \ge 0$, $k \in \natural$, let $\X^k_t$ be the weighted Sobolev 
space defined by the norm
$$
  \|u\|_{\X^0_t}^2 \,=\, \int_{-\infty}^0 \e^{-\gamma_- x\e^{t/2}}|u(x)|^2
   \d x + \int_0^\infty |u(x)|^2\d x~, \quad
  \|u\|_{\X^k_t}^2 \,=\, \sum_{i=0}^k \|\partial_x^i u\|_{\X^0_t}^2~.
$$
If $(u,v) \in \CC([t_0,t_1],\Z_t)$ is a solution of \equ(uv), it follows from
\clm(Fbound) and from the definition \equ(primitives) of $F,G$ that 
$(F,G) \in \X^2_t \times \X^1_t$ for all $t \in [t_0,t_1]$. Moreover, using 
a density argument as in the proof of \clm(alpha), one verifies that 
$(F,G) \in \CC^1([t_0,t_1],\X^1_t \times \X^0_t)$ is a classical solution of 
\equ(FG). As in Definition~1.1, this means that, if
$$
   \tilde F(\xi,t) \,=\, F(\xi \e^{-t/2},t)~, \quad
   \tilde G(\xi,t) \,=\, G(\xi \e^{-t/2},t)~,
$$
then $(\tilde F,\tilde G) \in \CC^1([t_0,t_1],\X^1_0 \times \X^0_0) \cap
\CC([t_0,t_1],\X^2_0 \times \X^1_0)$. For later use, we also note that
$$
   \tilde F_t(\xi,t) \,=\, \bigl(F_t - {x \over 2}F_x\bigr)(\xi\e^{-t/2},t)
   ~, \quad \tilde G_t(\xi,t) \,=\, \bigl(G_t - {x \over 2}G_x\bigr)
   (\xi\e^{-t/2},t)~. \EQ(tildeFG)
$$


\SECTION Energy Estimates

As in the previous section, we assume that $(u,v) \in \CC([t_0,t_1],\Z_t)$ 
is a solution of \equ(uv) satisfying the bound \equ(ubound). To control 
the time behavior of the functions $f,g$ defined in \equ(decomp), we shall
use five pairs of energy functionals. 

We first introduce unweighted functionals for the primitives $F,G$
defined in \equ(FG): 
$$
  E_0(t) \,=\, \int_\real \left({1 \over 2}F^2 + \eta \e^{-t} 
   FG\right)\d x~, \quad
  \EE_0(t) \,=\, {1 \over 2} \int_\real \left(F_x^2 + \eta \e^{-t}
   G^2\right)\d x~. \EQ(E0)
$$

\CLAIM Lemma(E0) Assume that $(u,v) \in \CC([t_0,t_1],\Z_t)$ is 
a solution of \equ(uv). Then $E_0$ and $\EE_0$ belong to $\CC^1([t_0,t_1])$ 
and
$$\eqalign{
  \dot E_0 \,=&\, -{E_0 \over 2} + \int_\real \left(-F_x^2 + {\e^t
   \over 2}\gamma'(x\e^{t/2})F^2 + \eta \e^{-t}G^2 -2\nu\e^{-t/2}F_xG
   + FR\right)\d x~, \cr
  \dot \EE_0 \,=&\, {\EE_0 \over 2} + \int_\real \left(-G^2 -\e^{t/2}
   \gamma(x\e^{t/2})F_x G + GR\right)~,}
$$
for all $t \in [t_0,t_1]$, where $R$ is defined in \equ(primR). 

\REMARK Here and in the sequel, we use the notation $\dot E = (\dd E/\dd t)$,
$\dot \EE = (\dd \EE/\dd t)$. 

\PROOF Since $(F,G) \in \CC^1([t_0,t_1],\H^1 \times \L^2)$, the functions
$E_0, \EE_0$ belong to $\CC^1([t_0,t_1])$, and a direct calculation yields:
$$\eqalign{
   \dot E_0(t) \,&=\, \int_\real \left(FF_t + \eta\e^{-t}\bigl((FG)_t 
    - FG\bigr)\right)\d x~, \cr 
   \dot \EE_0(t) \,&=\, \int_\real \left(-F_{xx}F_t + \eta\e^{-t}\bigl(
    GG_t - \frac{1}{2} G^2\bigr)\right)\d x~.}
$$
Using the identities
$$\eqalign{
   FF_t &+ \eta \e^{-t}\bigl((FG)_t -\frac{x}{2}(FG)_x -FG -G^2\bigr) \cr
   &= F F_{xx} + (\frac{x}{2} -\e^{t/2}\gamma(x\e^{t/2}))FF_x
   +2\nu\e^{-t/2}FG_x +FR~,} \EQ(identFG)
$$
$$\eqalign{
   -F_{xx}F_t &+ \eta \e^{-t}\bigl(GG_t -\frac{x}{2}GG_x -G^2\bigr) \cr
   &= -G^2 -\frac{x}{2}F_xF_{xx} -\e^{t/2}\gamma(x\e^{t/2}) GF_x
   + 2\nu\e^{-t/2}GG_x +GR~,} \EQ(identGG)
$$
which follow from \equ(FG), and integrating by parts, we obtain the 
desired expressions. This concludes the proof of \clm(E0). \QED

We next introduce weighted functionals for the primitives $F,G$:
$$\eqalign{
  E_1(t) \,&=\, \int_\real {\e^t \over p(x\e^{t/2})}  \left({1 \over
   2}F^2 + \eta \e^{-t} FG\right)\d x~, \cr
  \EE_1(t) \,&=\, {1 \over 2} \int_\real {\e^t \over p(x\e^{t/2})} 
   \left(F_x^2 + \eta \e^{-t} G^2\right)\d x~,} \EQ(E1)
$$
where the weight $p$ is defined in \equ(pdiff). 

\CLAIM Lemma(E1) Assume that $(u,v) \in \CC([t_0,t_1],\Z_t)$ is 
a solution of \equ(uv). Then $E_1$ and $\EE_1$ belong to $\CC^1([t_0,t_1])$ 
and
$$\eqalign{
  \dot E_1 \,=&\, {E_1 \over 2} + \int_\real {\e^t \over p(x\e^{t/2})}  
   \left(-F_x^2 +\eta \e^{-t}G^2 +2\nu\e^{-t/2}FG_x + FR\right)\d x~, \cr
  \dot \EE_1 \,=&\, {3\EE_1 \over 2} + \int_\real {\e^t \over 
   p(x\e^{t/2})} \left(-G^2 + 2\nu\e^{-t/2}GG_x + GR\right)\d x~,} 
$$
for all $t \in [t_0,t_1]$. 

\PROOF We remark that
$$
  E_1(t) \,=\, \int_\real {\e^{t/2} \over p(\xi)} \left({1 \over2}
   \tilde F^2 + \eta \e^{-t} \tilde F \tilde G\right)\d \xi~, \quad
  \EE_1(t) \,=\, {1 \over 2} \int_\real {\e^{3t/2} \over p(\xi)} 
   \left(\tilde F_\xi^2 + \eta \e^{-2t} \tilde G^2\right)\d \xi~,
$$
where $\tilde F(\xi,t) = F(\xi \e^{-t/2},t)$, $\tilde G(\xi,t) = 
G(\xi \e^{-t/2},t)$. Since $(\tilde F,\tilde G) \in \CC^1([t_0,t_1],\X^1_0 
\times \X^0_0)$ (see the remark at the end of the previous section), it
follows that $E_1, \EE_1 \in \CC^1([t_0,t_1])$. Using \equ(tildeFG), we
thus find
$$\eqalign{
  \dot E_1 \,=&\, {E_1 \over 2} + \int_\real {\e^t \over p(x\e^{t/2})}  
   \left(FF_t -\frac{x}{2}FF_x +\eta \e^{-t}\bigl((FG)_t -\frac{x}{2}(FG)_x
   -FG\bigr)\right)\d x~, \cr
  \dot \EE_1 \,=&\, {3\EE_1 \over 2} + \int_\real {\e^t \over p(x\e^{t/2})} 
  \left(F_x F_{xt} -\frac{x}{2}F_x F_{xx} -\frac{1}{2}F_x^2 +\eta 
   \e^{-t}\bigl(GG_t -\frac{x}{2} GG_x -G^2\bigr)\right)\d x~.} 
$$
Applying the identities \equ(identFG), \equ(identGG) and the relation
$F_t = G + {x \over 2}F_x$, we obtain the desired result after some 
integrations by parts. This concludes the proof of \clm(E1). \QED

We now define positive constants $A_0$, $B_0$ by
$$
   A_0 \,=\, 2\left(\inf_{\xi \ge 0} p(\xi)|\gamma'(\xi)|\right)^{-1}~,
   \quad B_0 \,=\, \left(\sup_{\xi \in \real}p(\xi)\gamma(\xi)^2\right)^{-1}
   ~. \EQ(A0B0)
$$
Due to \equ(gamasym), \equ(gamprim), \equ(pbound), these quantities are 
well-defined. Moreover, the inequality $|\gamma'(\xi)| \le {1 \over 2}
\gamma(\xi)^2$ implies that $A_0 \ge 4B_0 > 0$. With these notations, we 
introduce the functional
$$
   S_1(t) \,=\, A_0 E_0(t) + B_0 \EE_0(t) + 2E_1(t) + \EE_1(t)~, \quad
   t \in [t_0,t_1]~.
$$

\CLAIM Proposition(S1) Assume that $\eta\e^{-t_0}$ is sufficiently small, 
and that $(u,v) \in \CC([t_0,t_1],\Z_t)$ is a solution of \equ(uv) satisfying
the bound \equ(ubound). Then $S_1 \in \CC^1([t_0,t_1])$, $S_1(t) \ge 0$, 
and there exist positive constants $K_5$, $K_6$ such that, for all 
$t \in [t_0,t_1]$,
$$
  \dot S_1(t) + {1 \over 2} S_1(t) \,\le\, -K_5 \int_0^\infty (x^2{+}x^4)
  f^2 \d x + K_6\e^{-t/4}(\|f\|_{\L^2_t}+\|g\|_{\L^2_t})M(t)~, \EQ(S1)
$$
where $M(t)^2 \,=\, \alpha(t)^2 + \beta(t)^2 + \|f(t)\|_{\H^1_t}^2 + 
\|g(t)\|_{\L^2_t}^2$. 

\PROOF Assuming $\eta \e^{-t_0} \le \min(1/2, B_0/A_0)$, one verifies 
that $A_0 E_0(t) + B_0 \EE_0(t) \ge 0$ and $2E_1(t) + \EE_1(t) \ge 0$
for $t \in [t_0,t_1]$. Next, we remark that $F_x = \e^{-t}p(x\e^{t/2})f$,
hence $\|F_x\|_{\L^2} \le C\|f\|_{\L^2_t}$ by \equ(norms), \equ(pbound). 
Thus, using \clm(Fbound) and \clm(Rbound), we deduce from \clm(E0) that
$$
   \dot E_0(t) + {E_0(t) \over 2} \,\le\, \int_\real \left(-F_x^2 + 
   {\e^t \over 2}\gamma'(x\e^{t/2})F^2\right)\d x + C \e^{-t/4}
   (\|f\|_{\L^2_t} + \|g\|_{\L^2_t})M(t)~.
$$
Similarly, using the bound $|\e^{t/2}\gamma(x\e^{t/2})F_x G| \le 
{1 \over 2}(G^2 +\e^t\gamma(x\e^{t/2})^2 F_x^2)$, we obtain
$$
   \dot \EE_0(t) + {\EE_0(t) \over 2} \,\le\, {1 \over 2}\int_\real
   \left(-G^2 + F_x^2 + \e^t \gamma(x\e^{t/2})^2 F_x^2\right)\d x + C 
   \e^{-t/4} \|g\|_{\L^2_t}M(t)~.
$$
Finally, applying \clm(Fbound) and \clm(Rbound) again, we deduce from 
\clm(E1) that
$$\eqalign{
  2\dot E_1(t) &+\dot \EE_1(t) + E_1(t) + {1 \over 2}\EE_1(t) \,\le\, 
   C \e^{-t/4}(\|f\|_{\L^2_t} + \|g\|_{\L^2_t})M(t) \cr
  &+ \int_\real {\e^t \over p(x\e^{t/2})} \left(-F_x^2 -G^2 + F^2 
   +2\nu\e^{-t/2}(2F+G)G_x \right)\d x~.}
$$
The last term in the right-hand side is bounded with the help of 
\equ(primitives), \equ(pbound) and \clm(Fbound):
$$\eqalign{
   \int_\real {\e^{t/2} \over p(x\e^{t/2})}|(2F{+}G)G_x| \d x \,&\le\, 
    C \left(\int_\real {\e^t (F^2{+}G^2) \over p(x\e^{t/2})}\d x\right)^{1/2}
    \left(\int_\real \e^{-2t} p(x\e^{t/2}) g^2 \d x\right)^{1/2} \cr
   \,&\le\, C \e^{-t/2} (\|f\|_{\L^2_t} +\|g\|_{\L^2_t}) \|g\|_{\L^2_t}~.}
$$
Combining these estimates and using \equ(F1), \equ(A0B0) together with the
inequality $\gamma'(\xi) \le 0$ for $\xi \le 0$, we obtain
$$\eqalign{
  \dot S_1(t) + {S_1(t) \over 2} \,&\le\, -{B_0 \over 2} \int_\real 
   (7 F_x^2 + G^2)\d x - \int_\real {\e^t \over p(x\e^{t/2})}
   \left({1 \over 2}F_x^2 + G^2\right)\d x \cr
  &+ C \e^{-t/4} (\|f\|_{\L^2_t} + \|g\|_{\L^2_t})M(t)~,}
$$
for all $t \in [t_0,t_1]$, and \equ(S1) follows using \equ(primitives), 
\equ(pbound). This concludes the proof of \clm(S1). \QED

In the rest of this section, we introduce three pairs of weighted functionals
$E_i,\EE_i$ ($i=2,3,4$) to control the solutions $(f,g)$ of \equ(fg) in the 
space $\Z_t$. To each pair will correspond a different weight function $p_i : 
\real \to \real_+$. To define the weight $p_2$, we choose a smooth function 
$\chi_2 : \real \to (0,1]$ satisfying $\chi_2(\xi) = 2\kappa/\gamma_- < 1$ 
for $\xi \le -1$ and $\chi_2(\xi) = 1$ for $\xi \ge 0$. We set $\gamma_2 = 
\chi_2 \gamma$. The weight $p_2 : \real \to \real_+$ is then the (unique) 
solution of the differential problem
$$
   p_2'(\xi) \,=\, \gamma_2(\xi)p_2(\xi)~, \quad \xi \in \real~, 
   \quad \lim_{\xi \to +\infty} {p_2(\xi) \over \xi^2} \,=\, 1~. 
   \EQ(p2diff)
$$
Clearly, $p_2(\xi) = p(\xi)$ for $\xi \ge 0$, and there exists $C \ge 1$ 
such that $C^{-1}\e^{2\kappa\xi} \le p_2(\xi) \le C\e^{2\kappa\xi}$ for 
$\xi \le 0$. In particular, we have for all $u \in \L^2_t$
$$
   \int_{-\infty}^0 p_2(x\e^{t/2})u(x)^2\d x + \int_0^\infty \e^{-t} 
   p_2(x\e^{t/2})u(x)^2\d x \,\le\, C\|u\|_{\L^2_t}^2~. \EQ(L1)
$$
We now define the functionals
$$\eqalign{
  E_2(t) \,&=\, \int_\real \e^{-t} p_2(x\e^{t/2}) \left({1 \over
   2}f^2 + \eta \e^{-t} fg\right)\d x~, \cr
  \EE_2(t) \,&=\, {1 \over 2} \int_\real \e^{-t} p_2(x\e^{t/2}) 
   \left(f_x^2 + \eta \e^{-t} g^2 \right)\d x~,} \EQ(E2)
$$
together with $S_2(t) = 2E_2(t) + \EE_2(t)$. 

\CLAIM Proposition(S2) Assume that $\eta\e^{-t_0} \le 1/8$, and that
$(u,v) \in \CC([t_0,t_1],\Z_t)$ is a solution of \equ(uv) satisfying
the bound \equ(ubound). Then $S_2 \in \CC^1([t_0,t_1])$ and there exist
positive constants $K_7$, $K_8$ such that, for all $t \in [t_0,t_1]$, 
$$
   S_2(t) \,\ge\, {1 \over 4} \int_\real \e^{-t} p_2(x\e^{t/2})\bigl(f^2 + 
   f_x^2 + \eta \e^{-t}g^2\bigr)\d x ~, \EQ(S2low)
$$
and
$$\eqalign{
  \dot S_2 + {1 \over 2}S_2 \,\le\, &-K_7 \left(\int_0^\infty 
   x^2 \bigl(f_x^2 + g^2\bigr)\d x + \int_{-\infty}^0 
   \e^{2\kappa x\e^{t/2}} f^2 \d x\right) \cr
  &+K_8 \left(\int_0^\infty x^2 f^2\d x + \e^{-t/4}(\|f\|_{\L^2_t} +
   \|g\|_{\L^2_t})M(t)\right)~.} \EQ(dotS2)
$$

\PROOF Since $2\eta\e^{-t}|fg| \le 4\eta \e^{-t}f^2 + {1 \over 4}\eta
\e^{-t}g^2$ and $\eta \e^{-t} \le 1/8$, the lower bound \equ(S2low) 
is obvious. To compute the time derivative of $E_2$, we note that
$$
  E_2(t) \,=\, \int_\real \e^{-3t/2} p_2(\xi) \left({1 \over
   2}\tilde f^2 + \eta \e^{-t} \tilde f \tilde g\right)\d \xi~,
$$
where $\tilde f(\xi,t) = f(\xi \e^{-t/2},t)$, $\tilde g(\xi,t) = 
g(\xi \e^{-t/2},t)$. If we assume that $(u(t_0),v(t_0)) \in \H^2_{t_0} \times
\H^1_{t_0}$, then (as in \clm(localW)) $(\tilde f, \tilde g) \in 
\CC([t_0,t_1],\H^2_0 \times \H^1_0) \cap \CC^1([t_0,t_1],\H^1_0 \times 
\L^2_0)$ and $\tilde f_t(\xi,t) = (f_t - {x \over 2}f_x)(\xi\e^{-t/2},t)$,
$\tilde g_t(\xi,t) = (g_t - {x \over 2}g_x)(\xi\e^{-t/2},t)$. 
A direct calculation then yields
$$
   \dot E_2(t) \,=\, \int_\real \e^{-t} p_2(x\e^{t/2})\left(ff_t 
   -{x \over 2}ff_x -{3 \over 4}f^2 + \eta\e^{-t}\bigl((fg)_t - {x \over 2}
   (fg)_x -{5 \over 2}fg\bigr)\right)\d x~.
$$
Applying the identity
$$\eqalign{
   ff_t &+ \eta \e^{-t}\bigl((fg)_t -\frac{x}{2}(fg)_x -4fg -g^2\bigr) \cr
   &= ff_{xx} + (\frac{x}{2} +\e^{t/2}\gamma(x\e^{t/2}))ff_x
   +\frac{3}{2} f^2 + \nu\gamma(x\e^{t/2})fg + 2\nu\e^{-t/2}fg_x
   +fr~,} \EQ(identfg)
$$
which follows from \equ(fg), and integrating by parts, we obtain
$$\eqalign{
  \dot E_2 \,=\, {3E_2 \over 2} + \int_\real \e^{-t} p_2(x\e^{t/2})  
  &\Bigl(-f_x^2 +\frac{1}{2}\e^t\Gamma_2(x\e^{t/2})f^2 +\eta 
   \e^{-t}g^2 \cr 
  &+\nu(\gamma-2\gamma_2)(x\e^{t/2})fg -2\nu\e^{-t/2}f_xg + fr\Bigr)\d x~,}
  \EQ(dotE2)
$$
where $\Gamma_2 = \gamma_2'-\gamma'-\gamma_2(\gamma-\gamma_2)$. As is 
easily verified, the right-hand side of \equ(dotE2) is a continuous 
function of the initial data $(u(t_0),v(t_0))$ in the topology of $\Z_{t_0}$, 
uniformly in $t \in [t_0,t_1]$. Therefore, using a density argument as
in the proof of \clm(alpha), we conclude that $E_2 \in \CC^1([t_0,t_1])$
and that \equ(dotE2) holds in the general case where $(u(t_0),v(t_0)) \in
\Z_{t_0}$ only. 

In a similar way, we obtain for regular data
$$
  \dot \EE_2 \,=\, \int_\real \e^{-t} p_2(x\e^{t/2})\Bigl( f_xf_{xt}-
  {x \over 2}f_x f_{xx} -{3 \over 4} f_x^2 +\eta\e^{-t}\bigl(gg_t 
  -{x \over 2}gg_x -{5 \over 4}g^2\bigr)\Bigr)\d x~.
$$
Using the relation $f_t = g + {x \over 2}f_x + {3 \over 2}f$ as well as
the identity
$$\eqalign{
   -f_{xx}f_t &+ \eta \e^{-t}\bigl(gg_t -\frac{x}{2}gg_x -\frac{5}{2}
    g^2\bigr) \,=\, -(1-\nu\gamma(x\e^{t/2}))g^2 + 2\nu\e^{-t/2}gg_x\cr
   &-\frac{x}{2} f_xf_{xx} -\frac{3}{2}ff_{xx}
   +\e^{t/2}\gamma(x\e^{t/2}) f_x g +gr~,} \EQ(identgg)
$$
which follows from \equ(fg), we obtain after integrating by parts
$$
  \dot \EE_2 \,=\, {5\EE_2 \over 2} +\int_\real \e^{-t} p_2(x\e^{t/2})  
   \Bigl(-g^2 + \bigl(\nu g^2 +\e^{t/2} f_x g\bigr)(\gamma-\gamma_2)
   (x\e^{t/2}) +gr\Bigr)\d x~. \EQ(dotEE2)
$$
By the same density argument, $\EE_2 \in \CC^1([t_0,t_1])$ and \equ(dotEE2)
holds for all solutions $(u,v)$ of \equ(uv) in $\Z_t$. 

We now estimate the right-hand side of \equ(dotE2). Since $|(\gamma 
-2\gamma_2)(\xi)| \le \gamma(\xi)$ for $\xi \in \real$ and $\e^{-t}
p_2(x\e^{t/2})\gamma(x\e^{t/2}) \le C\e^{-t/2}(1{+}x)$ for $x \ge 0$, we 
obtain with the help of \equ(L1)
$$
   \int_\real \e^{-t}p_2(x\e^{t/2})|(\gamma-2\gamma_2)(x\e^{t/2})fg|\d x
   \,\le\, C\e^{-t/2}\|f\|_{\L^2_t} \|g\|_{\L^2_t}~. \EQ(L2)
$$
Remarking that $\Gamma_2(\xi) = 0$ for $\xi \ge 0$, we deduce from 
\equ(L1), \equ(dotE2), \equ(L2) and \clm(rmbound) that
$$\eqalign{
  \dot E_2 + {1 \over 2}E_2 \,&\le\, \int_0^\infty \e^{-t}p_2(x\e^{t/2})
   (f^2-f_x^2)\d x + {1 \over 2}\int_{-\infty}^0 p_2(x\e^{t/2}) 
   \Gamma_2(x\e^{t/2})f^2\d x \cr
  &+ C \e^{-t/4}(\|f\|_{\L^2_t} +\|g\|_{\L^2_t})M(t)~.} \EQ(dotE2bis)
$$
Since $\Gamma_2(\xi) \to -2\kappa(\gamma_- -2\kappa) = -2\kappa c_*$ as 
$\xi \to -\infty$, we can write $\Gamma_2(\xi) \le -\kappa c_* + 
\tilde \Gamma_2(\xi)$ for all $\xi \le 0$, where the support of
$\tilde \Gamma_2$ is contained in a compact interval $[-A,0]$. Applying 
\clm(Linfty), we thus obtain
$$\eqalign{
  \int_{-\infty}^0 p_2(x\e^{t/2}) &\Gamma_2(x\e^{t/2})f^2 \d x 
  + \kappa c_* \int_{-\infty}^0 p_2(x\e^{t/2})f^2\d x \cr
   \,&\le\, \e^{-t/2}\sup_{x\ge -A\e^{-t/2}}|f(x,t)|^2 \int_{-A}^0 
   p_2(\xi)\tilde \Gamma_2(\xi)\d \xi \,\le\, C \e^{-t/2} 
   \|f\|_{\H^1_t}^2~.} \EQ(Gamma2)
$$
Similarly, remarking that $\gamma_2(\xi) = \gamma(\xi)$ for $\xi \ge 0$, 
we deduce from \equ(L1), \equ(dotEE2) and \clm(rmbound) that
$$
  \dot \EE_2 + {1 \over 2}\EE_2 \,\le\, \int_0^\infty \e^{-t}p_2(x\e^{t/2})
   (\frac{3}{2}f_x^2 -g^2)\d x + C \e^{-t/4}(\|f\|_{\L^2_t} 
  +\|g\|_{\L^2_t})M(t)~. \EQ(dotEE2bis)
$$ 
Combining \equ(dotE2bis), \equ(Gamma2), \equ(dotEE2bis) and using 
\equ(pbound), \equ(p2diff), we obtain \equ(dotS2). This concludes the proof 
of \clm(S2). \QED

The construction of our next functionals $E_3$, $\EE_3$ is one of the main 
difficulties in the proof of \clm(main). The aim is to control the quantity
$$
   \int_{-\infty}^0 \e^{2\kappa x\e^{t/2}} \left(f^2 + f_x^2 +\eta\e^{-t}g^2
   \right)\d x + \int_0^\infty \left(f^2 + f_x^2 +\eta\e^{-t}g^2
   \right)\d x~,
$$
which is part of the norm of $(f,g)$ in $\Z_t$. A natural idea is to define
$E_3$, $\EE_3$ by the formulas \equ(E2) with $\e^{-t} p_2(x\e^{t/2})$ replaced
by $p_3(x\e^{t/2})$, where $p_3(\xi) = \OO(\e^{2\kappa\xi})$ as $\xi \to 
-\infty$ and $p_3(\xi) \to 1$ as $\xi \to +\infty$. However, we are not able 
to estimate properly the time derivative of these functionals without 
including in $\EE_3$ an additional term of the form
$$
   \int_\real p_3(x\e^{t/2})\lambda(x\e^{t/2})\gamma(x\e^{t/2})(\nu f_x^2
   -\eta\e^{-t/2}f_x g)\d x~,
$$
see (3.25) below. With this modification, the derivative of $\EE_3$ contains 
a quadratic form $Q(x,t)$ depending on the functions $\lambda$ and $p_3$, see
(3.30). 
As we shall show, the evolution of $E_3, \EE_3$ can then be controlled 
provided $Q(x,t)$ is positive definite. 

We now construct positive functions $\lambda, p_3$ so that the quadratic form 
$Q(x,t)$ in (3.30) 
is positive definite. First of all, since $\gamma_- = c_* +2\kappa > c_*$ and 
$\nu c_* < 1$ by \equ(etanugam), we can introduce
$$
   \lambda_- \,=\, \left({\gamma_-^2 \over c_*^2} -\nu\gamma_-\right)^{-1}
   \,>\, 0~. \EQ(lambda3)
$$
For later use, we remark that 
$$
   \lambda_- (1-\nu c_*) \,<\, (c_*/\gamma_-)^2 < 1~, \quad
   \hbox{and}\quad \lambda_- \gamma_- \,<\, \nu/\eta~. \EQ(plam)
$$
Next, in view of \equ(gamasym), \equ(gamprim), we can choose $\xi_3 > 0$ 
sufficiently large so that
$$
   \gamma(\xi_3) \,<\, c_* \lambda_-~, \quad \nu \gamma(\xi_3) \,\le\, 
   {1 \over 2}~, \quad \gamma'(\xi) \,\le\, -{1 \over 4}\gamma(\xi)^2 
   \quad \hbox{for all}\quad \xi \ge \xi_3~.
   \EQ(xi3)
$$
Remark that the first condition in \equ(xi3) is automatically satisfied if 
$\lambda_- \ge 1$, since $\gamma(0) = c_*$ and $\gamma$ is non-increasing. 
Now, let $\lambda : \real \to \real_+$ be a smooth, monotone function 
satisfying $\lambda(\xi) = \lambda_-$ if $\xi \le 0$, $\lambda(\xi) = 1$ if 
$\xi \ge \xi_3$, $(\lambda\gamma)'(\xi) \le 0$ for all $\xi \in \real$, and
$$
   \lambda(\xi)\bigl((1-\nu\gamma(\xi))^2 + \eta\gamma(\xi)^2\bigr) 
   \,\le\, 1~, \quad \xi \in [0,\xi_3]~. \EQ(lambound)
$$
Constructing such a function $\lambda$ is easy. Indeed, if $\lambda_- < 1$, 
the first condition in \equ(xi3) ensures that $\lambda$ can be chosen so that 
$(\lambda\gamma)'(\xi) \le 0$ for $\xi \in [0,\xi_3]$. On the other hand, we 
observe that the function
$$
   \Omega(\gamma) \,=\, (1-\nu\gamma)^2 + \eta\gamma^2 \,\equiv\,
   1 -2\nu\gamma + {\nu \gamma^2 \over c_*}
$$
is non-increasing for $\gamma \le c_*$, with $\Omega(0) = 1$ and 
$\Omega(c_*) = 1-\nu c_* > 0$. Since $\gamma(\xi) \le c_*$ for $\xi \ge 0$, 
the condition \equ(lambound) is obviously satisfied if $\lambda_- \le 1$.
If $\lambda_- > 1$, we remark that $\lambda_- \Omega(\gamma(0)) < 
(c_*/\gamma_-)^2 < 1$ by \equ(plam), hence it is sufficient to assume that 
$\lambda(\xi)$ decays rapidly enough to $1$ (as $\xi$ varies from $0$ to
$\xi_3$) so that \equ(lambound) is satisfied. 

We next define the weight function $p_3$. Let $\chi_3 : \real \to (0,1]$ 
be a smooth function satisfying $\chi_3(\xi) = 2\kappa/\gamma_- < 1$ 
for $\xi \le -1$, $\chi_3(\xi) = 1$ for $\xi \in [0,\xi_3]$, and 
$\chi_3(\xi) = 0$ for $\xi \ge \xi_3+1$. We also assume that
$\xi\chi_3'(\xi) \le 0$ for all $\xi \in \real$. We set $\gamma_3 = \chi_3 
\gamma$, and define the weight function $p_3 : \real \to \real_+$ as the 
(unique) solution of the differential problem
$$
   p_3'(\xi) \,=\, \gamma_3(\xi)p_3(\xi)~, \quad \xi \in \real~,
   \quad \lim_{\xi \to +\infty} p_3(\xi) \,=\, 1~. \EQ(p3diff)
$$
Clearly, there exists $C \ge 1$ such that $C^{-1} \le p_3(\xi) \le C$
for $\xi \ge 0$ and $C^{-1} \e^{2\kappa\xi} \le p_3(\xi) \le C 
\e^{2\kappa\xi}$ for $\xi \le 0$. 

With these definitions, we now introduce the functionals
$$\eqalign{
  E_3(t) \,&=\, \int_\real p_3(x\e^{t/2}) \left({1 \over 2}f^2 + 
   \eta \e^{-t} fg\right)\d x~, \cr
  \EE_3(t) \,&=\, {1 \over 2} \int_\real p_3(x\e^{t/2}) 
   \left(f_x^2 + \eta \e^{-t} g^2 + 2(\lambda\gamma)(x\e^{t/2})(\nu f_x^2
   -\eta\e^{-t/2}f_x g)\right)\d x~,} \EQ(E3)
$$
together with $S_3(t) = K E_3(t) + \EE_3(t)$, where $K = 3 + 4\nu 
\|\lambda\gamma\|_{\L^\infty}$. 

\CLAIM Proposition(S3) Assume that $\eta\e^{-t_0}$ is sufficiently small, 
and that $(u,v) \in \CC([t_0,t_1],\Z_t)$ is a solution of \equ(uv) satisfying
the bound \equ(ubound). Then $S_3 \in \CC^1([t_0,t_1])$, and there exist
positive constants $K_9$, $K_{10}$ such that, for all $t \in [t_0,t_1]$, 
$$
   S_3(t) \,\ge\, {1 \over 8} \int_\real p_3(x\e^{t/2})\bigl(f^2 + 
   f_x^2 + \eta \e^{-t}g^2\bigr)\d x~, \EQ(S3low)
$$
and
$$\eqalign{
  \dot S_3(t) + {1 \over 2}S_3(t) \,\le\, &-K_9 \int_\real p_3(x\e^{t/2})
   (g^2 +f_x^2 +\e^t \gamma(x\e^{t/2})^2 f_x^2)\d x \cr
  & +K_{10}\left(\int_{-\infty}^0 \e^{2\kappa x\e^{t/2}} f^2 \d x + 
   \int_0^\infty x^2 f_x^2 \d x\right) \cr
  & +K_{10} \left(\e^{-t/4}(\|f\|_{\H^1_t} + \|g\|_{\L^2_t})M(t) 
    +\e^{-t/2}M(t)^2\right)~.}
  \EQ(dotS3)
$$

\PROOF Since $|Kfg| \le {1 \over 8}g^2 + 2K^2 f^2$ and $|\e^{-t/2}\lambda
\gamma f_x g| \le {1 \over 4}\e^{-t}g^2 + (\lambda\gamma)^2 f_x^2$, we have
$$
  S_3 \,\ge\, \int_\real p_3(x\e^{t/2}) \Bigl(\bigl({K \over 2}-2K^2\eta
  \e^{-t}\bigr)f^2 + {1 \over 8}\eta \e^{-t} g^2 
  +{1 \over 2}f_x^2 \bigl(1+2\nu \lambda \gamma -
  2\eta\lambda^2 \gamma^2\bigr)(x\e^{t/2})\Bigr)\d x~.
$$
Assuming that $\eta\e^{-t_0} \le (8K)^{-1}$ and noting that $\nu - \eta
\lambda\gamma \ge \nu -\eta\lambda_-\gamma_- > 0$ by \equ(plam), 
we obtain \equ(S3low). 

Next, proceeding as in the proof of \clm(S2), we show that $E_3 \in 
\CC^1([t_0,t_1])$ and that 
$$\eqalign{
  \dot E_3 \,=\, {5E_3 \over 2} + \int_\real p_3(x\e^{t/2})  
  &\Bigl(-f_x^2 + \e^{t/2}(\gamma-\gamma_3)(x\e^{t/2})ff_x 
  +\eta \e^{-t}g^2 \cr 
  &+\nu(\gamma-2\gamma_3)(x\e^{t/2})fg -2\nu\e^{-t/2}f_xg + fr\Bigr)\d x~,}
 \EQ(dotE3)
$$
for $t \in [t_0,t_1]$. The analysis of $\EE_3$ is more complicated 
due to the additional term $2\lambda\gamma(\nu f_x^2 -\eta\e^{-t/2}f_x g)$.
First, assuming that the initial data are regular, we obtain by a direct
calculation
$$\eqalign{
  \dot \EE_3 \,&=\, \int_\real p_3(x\e^{t/2}) \Bigl((1+2\nu(\lambda
   \gamma)(x\e^{t/2}))\bigl(f_{x}f_{xt}-\frac{x}{2}f_x f_{xx} -\frac{1}{4}
   f_x^2\bigr) \cr
  &+ \eta\e^{-t}\bigl(gg_t -\frac{x}{2}gg_x -\frac{3}{4}g^2\bigr) 
   -\eta\e^{-t/2}(\lambda\gamma)(x\e^{t/2})\bigl((f_xg)_t -\frac{x}{2}
   (f_xg)_x -f_x g\bigr)\Bigr)\d x~.}
$$
Using the relation $f_t = g + {x \over 2}f_x + {3 \over 2}f$ together with
the identities \equ(identgg) and
$$\eqalign{
   2\nu f_x f_{xt} &- \eta \e^{-t/2}\bigl((f_xg)_t -\frac{x}{2}(f_xg)_x 
    -\frac{9}{2}f_x g -gg_x\bigr) \,=\, \e^{t/2}(1-\nu\gamma(x\e^{t/2}))
    f_x g \cr 
   &-\e^{t/2}f_xf_{xx} -\e^t\gamma(x\e^{t/2})f_x^2 +\nu xf_x
   f_{xx} + 4\nu f_x^2 -\e^{t/2}f_x r~,} \EQ(identfxg)
$$
which follow from \equ(fg), we obtain after integrating by parts
$$
  \dot \EE_3 \,=\, {7\EE_3 \over 2} +\int_\real p_3(x\e^{t/2})  
  \Bigl((g-\e^{t/2}(\lambda\gamma)(x\e^{t/2})f_x)r -Q(x,t)[\e^{t/2}
  \gamma(x\e^{t/2})f_x,g]\Bigr)\d x~, \EQ(dotEE3)
$$
where $Q(x,t)$ is the quadratic form defined by
$$\eqalign{
   Q(x,t)[z_1,z_2] \,=\, &z_1^2 \bigl(\lambda - \frac{1}{2}\lambda\gamma^{-1}
    (\gamma_3+\mu)\bigr)(x\e^{t/2}) -z_1 z_2 \bigl(1-\chi_3 + \lambda(1-\nu
    \gamma)\bigr)(x\e^{t/2}) \cr
   &+ z_2^2 \bigl(1+\nu(\gamma_3-\gamma) -\frac{1}{2}\eta\lambda\gamma
    (\gamma_3+\mu)\bigr)(x\e^{t/2})~, \quad (z_1, z_2) \in \real^2~,}
$$
and $\mu = (\lambda\gamma)'/(\lambda\gamma) \le 0$. By density, \equ(dotEE3)
holds for all solutions $(u,v)$ of \equ(uv) in $\Z_t$.

Applying \clm(rmbound) and recalling that $K = 3 + 4\nu 
\|\lambda\gamma\|_{\L^\infty}$, we deduce from \equ(dotE3), \equ(dotEE3) that
$$\eqalign{
  \dot S_3(t) &+ {1 \over 2}S_3(t) \,\le\, \int_\real p_3(x\e^{t/2}) 
   \Bigl(-f_x^2 +\frac{3}{2}K f^2 -Q(x,t)[\e^{t/2}\gamma(x\e^{t/2})f_x,g]\cr
  &-\e^{t/2} (\lambda\gamma)(x\e^{t/2})f_x r +\nu K (\gamma-2\gamma_3)
   (x\e^{t/2})fg + K\e^{t/2}(\gamma{-}\gamma_3)(x\e^{t/2})ff_x \Bigr)\d x\cr
  &+C \e^{-t/4} (\|f\|_{\H^1_t} + 
   \|g\|_{\L^2_t})M(t)~.} \EQ(dotS3first)
$$
We shall prove below that there exists $Q_0 > 0$ such that, for all 
$(x,t) \in \real \times \real_+$, 
$$
   Q(x,t)[z_1,z_2] \,\ge\, Q_0(z_1^2 + z_2^2)~, \quad (z_1, z_2) \in \real^2~.
   \EQ(Qpositive)
$$
Assuming for a while that \equ(Qpositive) holds, and using \clm(rmbound)
together with the inequalities
$$\eqalign{
   K\e^{t/2} \gamma(1-\chi_3)ff_x -\e^{t/2}\lambda\gamma f_x r \,&\le\, 
  {Q_0 \over 2}\e^t\gamma^2 f_x^2 + {K^2 \over Q_0}f^2 + {1 \over Q_0}
  \lambda^2 r^2~, \cr
  \nu K\gamma(1-2\chi_3)fg \,&\le\, {Q_0 \over 2}g^2 + {\nu^2 K^2 \over 
  2 Q_0}\gamma^2 f^2~,}
$$
we deduce from \equ(dotS3first) that
$$\eqalign{
  \dot S_3(t) &+ {1 \over 2}S_3(t) \,\le\, \int_\real p_3(x\e^{t/2})
    \Bigl(-f_x^2 -{Q_0 \over 2} (g^2 + \e^t \gamma(x\e^{t/2})^2 f_x^2)
    \Bigr)\d x\cr
  &+C \left(\int_\real p_3(x\e^{t/2}) f^2\d x + \e^{-t/2}M(t)^2 +
   \e^{-t/4} (\|f\|_{\H^1_t} + \|g\|_{\L^2_t})M(t)\right)~.} \EQ(dotS3sec)
$$
The estimate \equ(dotS3) is then a straightforward consequence of 
\equ(dotS3sec) and of the Hardy-type inequality
$$
  \int_\real p_3(x\e^{t/2}) f^2\d x \,\le\, C\left(\int_{-\infty}^0 
   \e^{2\kappa x\e^{t/2}} f^2 \d x + \int_0^\infty x^2 f_x^2 \d x\right)~.
$$

It remains to prove the property \equ(Qpositive), namely
$$
   \bigl(1-\chi_3 + \lambda(1-\nu \gamma)\bigr)^2 \,<\, 
   4 \bigl(\lambda - \frac{1}{2}\lambda\gamma^{-1} (\gamma_3+\mu)\bigr)
   \bigl(1+\nu(\gamma_3-\gamma) -\frac{1}{2}\eta\lambda\gamma 
   (\gamma_3+\mu)\bigr)~,
$$
for all $\xi \in [-\infty,+\infty]$. Expanding the products in both sides, 
we rewrite this condition in the equivalent form
$$\eqalign{
  (1-\chi_3)^2\bigl(1&+2\nu\lambda\gamma-\eta\lambda^2\gamma^2\bigr) 
   -2\lambda + \lambda^2 \bigl((1-\nu\gamma)^2 + \eta\gamma^2\bigr) \cr
  &< \eta\lambda^2 \mu^2 -2\lambda^2\eta \mu(\gamma-\gamma_3) -2\lambda
   \gamma^{-1}\mu(1-\nu(\gamma-\gamma_3))~.} \EQ(equiv)
$$
To prove \equ(equiv), we first remark that the right-hand side is positive, 
since $\mu \le 0$, $\gamma-\gamma_3 \ge 0$ and $1-\nu(\gamma-\gamma_3) 
\ge 1-\nu c_* > 0$. We also recall that $1 +2\nu\lambda\gamma - \eta\lambda^2
\gamma^2 \ge 1$, since $\nu -\eta\lambda\gamma >0$ by \equ(plam). 
We now distinguish three cases according to whether $\xi \le 0$, $\xi \in 
[0,\xi_3]$, or $\xi \ge \xi_3$. 

\noindent{\bf 1.} If $\xi \le 0$, then $\lambda = \lambda_-$ and $1-\chi_3 
\le c_*/\gamma_-$, hence it is sufficient to verify the stronger condition
$$
   {c_*^2 \over \gamma_-^2} \bigl(1 +2\nu\lambda_- \gamma -\eta\lambda_-^2
   \gamma^2\bigr) -2\lambda_- + \lambda_-^2 \bigl((1-\nu\gamma)^2 + 
   \eta\gamma^2\bigr) \,<\, 0~, \EQ(stronger)
$$
for all $\gamma \in [c_*,\gamma_-]$. Let $\Psi(\gamma)$ denote the 
left-hand side of \equ(stronger), considered as a function of $\gamma$. 
Using \equ(plam) and the relation $\nu^2 + \eta = \nu/c_*$, it is not 
difficult to verify that $\Psi$ is convex and that
$$
   \Psi(\gamma_-) \,=\, -\lambda_-^2 (1-\nu c_*)\left({\gamma_-^2 \over c_*^2}
   - 1\right) \,<\, 0~, \quad 
   \Psi'(c_*) \,=\, {2c_*^2 \lambda_- \over \gamma_-^2}\bigl(\nu 
   -\eta c_* \lambda_-\bigr) \,>\, 0~.
$$
Since $\Psi'' > 0$, it follows that $\Psi'(\gamma) \ge \Psi'(c_*) > 0$ for all 
$\gamma \ge c_*$, hence $\Psi(\gamma) \le \Psi(\gamma_-) < 0$ for all 
$\gamma \in [c_*,\gamma_-]$, which is the desired inequality. 

\noindent{\bf 2.} If $\xi \in [0,\xi_3]$, then $\chi_3 = 1$, hence 
the left-hand side of \equ(equiv) is negative by \equ(lambound). 

\noindent{\bf 3.} If $\xi \ge \xi_3$, then $\lambda = 1$, $1-\chi_3 \le 1$, 
hence the left-hand side of \equ(equiv) is bounded from above by $\nu^2
\gamma^2$. Neglecting the first two terms in the right-hand side (which
are positive) and noting that $\mu = \gamma'/\gamma \le 0$, we arrive at the 
stronger condition
$$
   \nu^2 \gamma(\xi)^2 \,\le\, -2 {\gamma'(\xi) \over \gamma(\xi)^2}
   \bigl(1-\nu\gamma(\xi)\bigr)~, \quad \xi \ge \xi_3~,
$$
which is satisfied by assumption on $\xi_3$, see \equ(xi3). This concludes
the proof of \clm(S3). \QED

Finally, we introduce our last functionals
$$\eqalign{
  E_4(t) \,&=\, \int_\real \e^{-3t} p_4(x\e^{t/2}) \left({1 \over
   2}f^2 + \eta \e^{-t} fg\right)\d x~, \cr
  \EE_4(t) \,&=\, {1 \over 2} \int_\real \e^{-3t} p_4(x\e^{t/2}) 
   \left(f_x^2 + \eta \e^{-t} g^2\right)\d x~,} \EQ(E4)
$$
where $p_4(\xi) = p(\xi)^3$. We set $S_4 = 2E_4 + \EE_4$. 

\CLAIM Proposition(S4) Assume that $\eta\e^{-t_0} \le 1/8$ and that
$(u,v) \in \CC([t_0,t_1],\Z_t)$ is a solution of \equ(uv) satisfying
the bound \equ(ubound). Then $S_4 \in \CC^1([t_0,t_1])$ and there exist
positive constants $K_{11}, K_{12}$ such that, for $t \in [t_0,t_1]$, 
$$
   S_4(t) \,\ge\, {1 \over 4}\int_\real \e^{-3t} p_4(x\e^{t/2})\bigl(f^2 + 
   f_x^2 + \eta \e^{-t}g^2\bigr)\d x~, \EQ(S4low) 
$$
and
$$\eqalign{
  \dot S_4(t) &+ {1 \over 2}S_4(t) \,\le\, -K_{11} \int_0^\infty x^6 
   (f_x^2 + g^2)\d x \cr
  &+ K_{12}\left(\int_0^\infty (x^4 f^2 + x^2 f_x^2)\d x + \e^{-t/4}
   (\|f\|_{\H^1_t} + \|g\|_{\L^2_t})M(t)\right)~.} \EQ(dotS4)
$$

\PROOF The lower bound \equ(S4low) is proved as in \equ(S2low). Arguing like 
in the preceding propositions, we show that $E_4, \EE_4 \in \CC^1([t_0,t_1])$ 
and that
$$\eqalign{
  \dot E_4 \,=\, -{E_4 \over 2} + \int_\real \e^{-3t}p_4(x\e^{t/2})  
  &\Bigl(-f_x^2 -2 \e^{t/2}\gamma(x\e^{t/2})ff_x +\eta \e^{-t}g^2 \cr 
  &-5\nu\gamma(x\e^{t/2})fg -2\nu\e^{-t/2}f_xg + fr\Bigr)\d x~,}
$$
$$\eqalign{
  \dot \EE_4 \,=&\, {\EE_4 \over 2} +\int_\real \e^{-3t} p_4(x\e^{t/2})  
   \Bigl(-g^2 -2\gamma(x\e^{t/2})(\nu g^2 +\e^{t/2}f_x g) + 
   gr\Bigr)\d x~.}
$$
Proceeding as in \equ(L2) and applying \clm(rmbound), we deduce that, for 
$t \in [t_0,t_1]$, 
$$\eqalign{
   \dot S_4(t) &+ {1 \over 2}S_4(t) \,\le\, C \e^{-t/4} (\|f\|_{\H^1_t} 
    +\|g\|_{\L^2_t})M(t) \cr
   &+\int_0^\infty \e^{-3t} p_4(x\e^{t/2}) \Bigl(-\frac{3}{2}f_x^2 -g^2 
   -2\e^{t/2}\gamma(x\e^{t/2})(f_x g + 2ff_x)\Bigr)\d x~.} \EQ(dotS4first)
$$
Since $p_4(\xi) = p(\xi)^3 \le C_0^3 (1+\xi)^6$ for $\xi \ge 0$, we have
for $x \ge 0$
$$\eqalign{
  \e^{-3t}(\gamma p_4)&(x\e^{t/2}) |2\e^{t/2}f_x g + 4\e^{t/2}ff_x| \cr 
  &\le \e^{-3t}p_4(x\e^{t/2})\bigl(\frac{1}{2}g^2 +f_x^2 + 
   2\e^{2t}\gamma(x\e^{t/2})^4f_x^2 + 8\e^t\gamma(x\e^{t/2})^2 f^2\bigr)\cr
  &\le \e^{-3t}p_4(x\e^{t/2})\bigl(\frac{1}{2}g^2 +f_x^2\bigr)
   +C \left(\e^{-t}(1{+}x\e^{t/2})^2 f_x^2 + \e^{-2t}(1{+}x\e^{t/2})^4 f^2
   \right)~,}
$$
for $x \ge 0$, and the estimate \equ(dotS4) follows from \equ(dotS4first). 
This concludes the proof of \clm(S4). \QED

We now summarize the decay properties of the four auxiliary functionals
$S_1,S_2,S_3$ and $S_4$. To this end, we define
$$
   S_5(t) \,=\, B_1 S_1(t) +B_2 S_2(t) +S_3(t) +S_4(t) + {1 \over 2}\eta 
   \e^{-t}\beta(t)^2~,
$$
where $B_2 = 1+K_7^{-1}(K_{10}+K_{12})$ and $B_1 = 1+K_5^{-1}(K_8B_2+K_{12})$.
In the proof of \clm(main), we shall use the following properties of
$S_5(t)$:
    
\CLAIM Proposition(S5) There exist constants $A_1,A_3,A_4 > 0$ and $A_2 \ge 1$
such that, if $\eta\e^{-t_0} \le A_1$ and if $(u,v) \in \CC([t_0,t_1],\Z_t)$ 
is a solution of \equ(uv) satisfying the bound \equ(ubound), then, for all 
$t \in [t_0,t_1]$,
$$
   A_2^{-1}S_5(t) \,\le\, \|f(t)\|_{\H^1_t}^2 + \eta\e^{-t}\left(
   \beta(t)^2 + \|g(t)\|_{\L^2_t}^2 \right) \,\le\, A_2 S_5(t)~,
   \EQ(bdS5)
$$
and
$$\eqalign{
  \dot S_5(t) + {1 \over 2}S_5(t) \,\le\, &-A_3 \left(\beta(t)^2 
   +\|g(t)\|_{\L^2_t}^2 + \|f_x(t)\|_{\L^2_t}^2\right) \cr
  &+ A_4 \e^{-t/4}\left(\|f(t)\|_{\H^1_t} +\|g(t)\|_{\L^2_t} +\e^{-t/4}M(t)
   \right)M(t)~,} \EQ(dotS5)
$$
where $M(t)^2 = \alpha(t)^2 + \beta(t)^2 + \|f(t)\|_{\H^1_t}^2 + 
\|g(t)\|_{\L^2_t}^2$. 

\PROOF Since $S_1(t) \ge 0$ by \clm(S1), the lower bound on $S_5$ in 
\equ(bdS5) follows immediately from \equ(S2low), \equ(S3low), \equ(S4low) 
and the properties of the weights $p_2,p_3,p_4$. The upper bound is proved 
in a similar way, using in addition \clm(Fbound) applied to $F$ and $G$. On 
the other hand, we have by \clm(alpha) and \clm(rmbound)
$$\eqalign{
  {\dd \over \dd t}\left({1 \over 2}\eta\e^{-t}\beta(t)^2\right)
   + {1 \over 4}\eta\e^{-t}\beta(t)^2 \,&=\, -\beta(t)^2 + {3 \over 4}
   \eta\e^{-t}\beta(t)^2 + m(t)\beta(t) \cr
  \,&\le\, -\beta(t)^2 + C \e^{-t/2} M(t)^2~.} \EQ(dotbeta)
$$
Combining the estimates \equ(S1), \equ(dotS2), \equ(dotS3), \equ(dotS4)
and \equ(dotbeta), we thus obtain
$$\eqalign{
  \dot S_5(t) &+ {1 \over 2}S_5(t) \,\le\, -K_5 \int_0^\infty (x^2+x^4)
   f^2 \d x -K_7 \int_{-\infty}^0 \e^{2\kappa x\e^{t/2}}f^2 \d x \cr
  &-\int_0^\infty (K_7 x^2 + K_{11}x^6)(f_x^2 + g^2)\d x -K_9 \int_\real 
   p_3(x\e^{t/2}) (g^2 + f_x^2)\d x -\beta(t)^2 \cr
  &+C\e^{-t/4} \left(\|f(t)\|_{\H^1_t} +\|g(t)\|_{\L^2_t} +\e^{-t/4}M(t)
   \right)M(t)~,}
$$
{}from which \equ(dotS5) follows using the properties of the weight $p_3$. 
This concludes the proof of \clm(S5). \QED

A useful consequence of \clm(S5) is:

\CLAIM Corollary(globound) There exist constants $A_5 > 0$ and $A_6 \ge 1$
such that, if $t_0 \ge A_5$ and if $(u,v) \in \CC([t_0,t_1],\Z_t)$ 
is a solution of \equ(uv) satisfying the bound \equ(ubound), then
$$
   \Phi_\eta(t,u(t),v(t)) \,\equiv\, \|u(t)\|_{\H^1_t}^2 +\eta\e^{-t}
   \|v(t)\|_{\L^2_t}^2 \,\le\, A_6 \Phi_\eta(t_0,u(t_0),v(t_0))~, 
   \EQ(globound)
$$
for all $t \in [t_0,t_1]$. 

\PROOF We introduce our last functional
$$
   S_6(t) \,=\, {1 \over 2}\alpha(t)^2 + \eta\e^{-t}\alpha(t)\beta(t) 
   +S_5(t)~, \quad t \in [t_0,t_1]~.
$$
In view of \equ(bdS5), if $\eta\e^{-t_0} \le \min(A_1,A_2^{-1})$, there
exists a constant $\C_1 \ge 1$ such that , for $t \in [t_0,t_1]$,
$$
   \C_1^{-1}S_6(t) \,\le\, \alpha(t)^2 + \|f(t)\|_{\H^1_t}^2 +\eta\e^{-t}
   \left(\beta(t)^2 + \|g(t)\|_{\L^2_t}^2 \right) \,\le\, \C_1 S_6(t)~.
   \EQ(bdS6)
$$
By \clm(equiv), il follows that
$$
   \C_2^{-1}S_6(t) \,\le\, \Phi_\eta(t,u(t),v(t)) \,\le\, \C_2 S_6(t)~, 
   \quad t \in [t_0,t_1]~, \EQ(S6Phi)
$$
for some $\C_2 \ge 1$. Now, since $\dot S_6(t) = \alpha(t)m(t) + \eta\e^{-t}
\beta(t)^2 + \dot S_5(t)$ by \equ(dotalpha), we deduce from \equ(rmbound) and
\equ(dotS5) that
$$
   \dot S_6(t) \,\le\, -A_3(\beta(t)^2 + \|g(t)\|_{\L^2_t}^2) 
   + \C_3 \e^{-t/4}M(t)^2~, \quad t \in [t_0,t_1]~,
$$
for some $\C_3 > 0$. Assuming that $\C_3 \e^{-t_0/4} \le A_3$ and using 
\equ(bdS6), we thus find $\dot S_6(t) \le \C_1 \C_3 \e^{-t/4}S_6(t)$ for 
$t \in [t_0,t_1]$, hence $S_6(t) \le \C_4 S_6(t_0)$ for $t \in [t_0,t_1]$, 
where $\C_4 = \exp(4\C_1 \C_3)$. Combining this estimate with \equ(S6Phi), 
we obtain \equ(globound). This concludes the proof of \clm(globound). \QED


\SECTION End of the Proof of \clm(main)

Let $t_0 \ge A_5$ and $\delta_0 \le (2A_6)^{-1/2}$, where $A_5, A_6$ are
as in \clm(globound). If $(u_0,v_0) \in \Z_{t_0}$ satisfies 
$\Phi_\eta(t_0,u_0,v_0) \le \delta_0^2$, then the system \equ(uv) has a
unique global solution $(u,v) \in \CC([t_0,+\infty),\Z_t)$ with 
$(u(t_0),v(t_0))= (u_0,v_0)$. Indeed, the local existence and uniqueness 
follow from \clm(localuv), and \clm(globound) shows that 
$\Phi_\eta(t,u(t),v(t)) \le 1/2$ as long as the solution $(u(t),v(t))$ 
exists. Then \clm(localuv), with $\delta_1 = 1/\sqrt{2}$, implies that the 
solution $(u(t),v(t))$ is globally defined. 

It remains to prove the decay estimate \equ(mainuv). Since
$$\eqalign{
  A_4 \e^{-t/4} \|g\|_{\L^2_t} M \,&\le\, {A_3 \over 4} \|g\|_{\L^2_t}^2 
   +C\e^{-t/2}M^2~, \cr
  A_4 \e^{-t/4} \|f\|_{\H^1_t} M \,&\le\, {A_3 \over 4} (\beta^2
   +\|g\|_{\L^2_t}^2) + A_4 \e^{-t/4}\|f\|_{\H^1_t}(|\alpha| +\|f\|_{\H^1_t})
   +C\e^{-t/2}M^2~,}
$$
it follows from \equ(dotS5) that
$$
   \dot S_5 + {1 \over 2} S_5 \,\le\, -{A_3 \over 2} (\beta^2
   +\|g\|_{\L^2_t}^2) + A_4 \e^{-t/4}\|f\|_{\H^1_t}(|\alpha| +\|f\|_{\H^1_t})
   +\C_5 \e^{-t/2}M^2~,
$$
for some $\C_5 > 0$. Setting $\rho_0^2 = \C_1 \C_2 A_6 \Phi_\eta(t_0,u_0,v_0)$,
we have $\alpha(t)^2 + \|f(t)\|_{\H^1_t}^2 \le \rho_0^2$ by \equ(globound),
\equ(bdS6), \equ(S6Phi), and $\|f(t)\|_{\H^1_t}^2 \le A_2 S_5(t)$ by 
\equ(bdS5). Therefore, assuming that $\C_5 \e^{-t_0/4} \le A_3/4$, we find
$$
   \dot S_5 +{1 \over 2}S_5 \,\le\, -{A_3 \over 4}(\beta^2 + 
   \|g\|_{\L^2_t}^2) + \C_6 \rho_0 \e^{-t/4} S_5^{1/2} + \C_5 \rho_0^2 
   \e^{-t/2}~, \quad t \ge t_0~,
$$
for some $\C_6 > 0$. Integrating this differential inequality and using the
bound $S_5(t_0) \le A_2 \rho_0^2$, we obtain after a short computation
$$
   S_5(t) + \int_{t_0}^t \e^{-(t-s)/2} (\beta(s)^2 + \|g(s)\|_{\L^2_s}^2)
   \d s \,\le \, \C_7 \rho_0^2 (1+(t{-}t_0)^2) \e^{-(t-t_0)/2}~,
   \EQ(prebound)
$$
for $t \ge t_0$, where $\C_7 > 0$ is independent of $t_0$ and $\rho_0$. 
In view of \equ(bdS5), this implies in particular
$$
   \|f(t)\|_{\H^1_t}^2 + \eta\e^{-t} (\beta(t)^2 + \|g(t)\|_{\L^2_t}^2)
   \,\le\,  A_2 \C_7 \rho_0^2 (1+(t{-}t_0)^2) \e^{-(t-t_0)/2},
   \quad t \ge t_0~. \EQ(bornefgb)
$$
Since $\dot \alpha(t) = \beta(t)$, we also deduce from \equ(prebound),
by a simple argument, that $\alpha(t)$ converges to some real number 
$\alpha^*$ as $t \to +\infty$, and that
$$
  |\alpha(t)-\alpha^*|^2 + \int_{t_0}^t \e^{-(t-s)/2} |\alpha(s) -
  \alpha^*|^2\d s \,\le\, \C_8  \rho_0^2 (1+(t{-}t_0)^2)\e^{-(t-t_0)/2}~,
   \quad t \ge t_0~, \EQ(alphadecay)
$$
for some $\C_8 > 0$. Finally, it follows from \equ(decomp) that
$$\eqalign{
  \|u(t)-\alpha^*\phi^*\|_{\H^1_t} \,&\le\, \|f(t)\|_{\H^1_t}
   + |\alpha(t)-\alpha^*| \|\phi(t)\|_{\H^1_t} + |\alpha^*|
   \|\phi(t) -\phi^*\|_{\H^1_t}~, \cr
  \|v(t)-\alpha^* \psi^*\|_{\L^2_t} \,&\le\, \|g(t)\|_{\L^2_t}+ |\beta(t)|
   \|\phi\|_{\L^2_t} + |\alpha(t)-\alpha^*|\|\psi\|_{\L^2_t}
    + |\alpha^*|  \|\psi(t) -\psi^*\|_{\L^2_t} ~,}
$$
hence the estimate \equ(mainuv) is a direct consequence of \equ(phipsibd),
\equ(phipsilim), \equ(prebound), \equ(bornefgb) and \equ(alphadecay). This 
concludes the proof of \clm(main). \QED


\REFERENCES

\input refs.tex
\end

%% file: header.tex
%
%
\magnification \magstep1
\hsize=16truecm
\vsize=22truecm
\voffset=0.9truecm
\hoffset=-0.1truecm
\normalbaselineskip=5.25mm
\baselineskip=5.25mm
\parskip=5pt
\parindent=20pt
\nopagenumbers
\headline={\ifnum\pageno>1 {\hss\tenrm-\ \folio\ -\hss} \else {\hfill}\fi}
\def\em{\sl}
%
%
\newcount\EQNcount \EQNcount=1
\newcount\CLAIMcount \CLAIMcount=1
\newcount\SECTIONcount \SECTIONcount=0
\newcount\SUBSECTIONcount \SUBSECTIONcount=0
\def\actualnumber{\number\SECTIONcount}
\newcount\timecount
\def\TODAY{\number\day~\ifcase\month\or January \or February \or March \or
  April \or May \or June
  \or July \or August \or September \or October \or November \or December \fi
  \number\year\timecount=\number\time
  \divide\timecount by 60}
\newdimen\strutdepth
\def\DRAFT{\def\lmargin(##1){\strut\vadjust{\kern-\strutdepth
  \vtop to \strutdepth{
  \baselineskip\strutdepth\vss\rlap{\kern-1.2 truecm\eightpoint{##1}}}}}
  \font\footfont=cmti7
  \footline={{\footfont \hfil File:\jobname, \TODAY,  \number\timecount h}}}
\def\lmargin(#1){}
\def\ifundefined#1{\expandafter\ifx\csname#1\endcsname\relax}
\def\ifff(#1,#2,#3){\ifundefined{#1#2}
  \expandafter\xdef\csname #1#2\endcsname{#3}\else
  \write16{Warning : doubly defined #1,#2}\fi}
\def\NEWDEF #1,#2,#3 {\ifff({#1},{#2},{#3})}
\def\EQ(#1){\lmargin(#1)\eqno\tag(#1)}
\def\NR(#1){&\lmargin(#1)\tag(#1)\cr}  
\def\tag(#1){({\rm \actualnumber.\number\EQNcount})
  \NEWDEF e,#1,(\actualnumber.\number\EQNcount)
  \global\advance\EQNcount by 1
  }
\def\equ(#1){\ifundefined{e#1}$\spadesuit$#1\else\csname e#1\endcsname\fi}
\def\CLAIM #1(#2) #3\par{
  \vskip.1in\medbreak\noindent
  {\lmargin(#2)\bf #1~\actualnumber.\number\CLAIMcount.} {\sl #3}\par
  \NEWDEF c,#2,{#1~\actualnumber.\number\CLAIMcount}
  \global\advance\CLAIMcount by 1
  \ifdim\lastskip<\medskipamount
  \removelastskip\penalty55\medskip\fi}
\def\CLAIMNONR #1 #2\par{
  \vskip.1in\medbreak\noindent
  {\bf #1.} {\sl #2}\par
  \ifdim\lastskip<\medskipamount
  \removelastskip\penalty55\medskip\fi}
\def\clm(#1){\ifundefined{c#1}$\spadesuit$#1\else\csname c#1\endcsname\fi}
\def\sectionsize{\twelvepoint}
\def\sectiontype{\bf}
\newskip\beforesectionskipamount  
\newskip\sectionskipamount        
\beforesectionskipamount=24pt plus8pt minus8pt
\sectionskipamount=3pt plus1pt minus1pt
\newskip\beforesubsectionskipamount  
\newskip\subsectionskipamount        
\beforesubsectionskipamount=12pt plus4pt minus4pt
\subsectionskipamount=2pt plus1pt minus1pt
\def\sectionskip{\vskip\sectionskipamount}
\def\beforesectionskip{\vskip\beforesectionskipamount}
\def\subsectionskip{\vskip\subsectionskipamount}
\def\beforesubsectionskip{\vskip\beforesubsectionskipamount}
\def\SECTION#1\par{\vskip0pt plus.3\vsize\penalty-75
  \vskip0pt plus -.3\vsize
  \global\advance\SECTIONcount by 1
  \def\actualnumber{\number\SECTIONcount}
  \beforesectionskip\noindent
  {\sectionsize\sectiontype \actualnumber.\ #1}
  \EQNcount=1
  \CLAIMcount=1
  \SUBSECTIONcount=0
  \nobreak\sectionskip\noindent}
\def\SECTIONNONR#1\par{\vskip0pt plus.3\vsize\penalty-75
  \vskip0pt plus -.3\vsize
  \global\advance\SECTIONcount by 1
  \beforesectionskip\noindent
  {\sectionsize\sectiontype #1}
  \EQNcount=1
  \CLAIMcount=1
  \SUBSECTIONcount=0
  \nobreak\sectionskip\noindent}
\def\SECT(#1)#2\par{\lmargin(#1)
  \SECTION #2\par
  \NEWDEF s,#1,{\actualnumber}}
\def\sec(#1){\ifundefined{s#1}$\spadesuit$#1
  \else Section \csname s#1\endcsname\fi}
\def\subsectionsize{}
\def\subsectiontype{\bf}
\def\SUBSECTION#1\par{\vskip0pt plus.2\vsize\penalty-75
  \vskip0pt plus -.2\vsize
  \global\advance\SUBSECTIONcount by 1
  \beforesubsectionskip\noindent
  {\subsectionsize\subsectiontype \actualnumber.\number\SUBSECTIONcount.\ #1}
  \nobreak\subsectionskip\noindent}
\def\SUBSECT(#1)#2\par{\lmargin(#1)
  \SUBSECTION #2\par
  \NEWDEF p,#1,{\actualnumber.\number\SUBSECTIONcount}}
\def\subsec(#1){\ifundefined{p#1}$\spadesuit$#1
  \else Subsection \csname p#1\endcsname\fi}
\def\pap{\rm}
\def\bok{\sl}
\newcount\BIBflag \BIBflag=0
\def\Item#1#2#3\par{%
  \ifnum\BIBflag=0 \NEWDEF b,#1,#2 \else \item{[#2]}\lmargin({#1})#3\par\fi
  \ifundefined{q#1}
  \ifnum\BIBflag=1
   \write16{Warning : unquoted reference [#2]}
  \fi\fi}
\def\refnb#1{%
  \ifundefined{b#1}$\spadesuit$#1\else \csname b#1\endcsname\fi
  \ifundefined{q#1}\expandafter\xdef\csname q#1\endcsname{}\fi}
\def\ref#1{[\refnb{#1}]}
\def\REFERENCES{\BIBflag=1
  \parindent=30pt
  \parskip=3pt  
  \SECTIONNONR References\par}

\def\APPENDIX(#1)#2\par{
  \def\actualnumber{#1}
  \SECTIONNONR Appendix #1. #2\par}
\def\PROOF{\medskip\noindent{\bf Proof.\ }}
\def\REMARK{\medskip\noindent{\bf Remark.\ }}
\def\REMARKS{\medskip\noindent{\bf Remarks.\ }}

\def\LIKEREMARK(#1){\medskip\noindent{\bf #1.}}
%
%
\let\endarg=\par
\def\finish{\def\endarg{\par\endgroup}}
\def\start{\endarg\begingroup}
\def\getNORMAL#1{{#1}}
\def\titlesize{\twelvepoint}
\def\titletype{\bf}
\def\TITLE{\beginTITLE\getTITLE}
  \def\beginTITLE{\start
     \titlesize\titletype\baselineskip=1.728
     \normalbaselineskip\rightskip=0pt plus1fil
     \noindent
     \def\endarg{\par\vskip.35in\endgroup}}
  \def\getTITLE{\getNORMAL}
\def\ENDTITLE{\endarg}
\def\AUTHOR{\beginAUTHOR\getAUTHOR}
  \def\beginAUTHOR{\start
    \vskip .25in\rm\noindent\finish}
  \def\getAUTHOR{\getNORMAL}
\def\FROM{\beginFROM\getFROM}
  \def\beginFROM{\start\parskip=0pt\vskip\baselineskip
    \def\finish{\def\endarg{\egroup\par\endgroup}}
    \vbox\bgroup\obeylines\eightpoint\em\finish}
  \def\getFROM{\getNORMAL}
\def\ABSTRACT#1\par{
  \vskip 1in {\noindent\sectionsize\sectiontype Abstract.} #1 \par}
\def\ENDABSTRACT{\vfill\break}
%
%
\newdimen\texpscorrection
\texpscorrection=0truecm  
\newcount\FIGUREcount \FIGUREcount=0
\newskip\ttglue 
\newdimen\figcenter
\def\figure #1 #2 #3 #4\cr{\null
  \global\advance\FIGUREcount by 1
  \NEWDEF fig,#1,{Fig.~\number\FIGUREcount}
  \write16{ FIG \number\FIGUREcount: #1}
  {\goodbreak\figcenter=\hsize\relax
  \advance\figcenter by -#3truecm
  \divide\figcenter by 2
  \midinsert\vskip #2truecm\noindent\hskip\figcenter
  \includegraphics{#1}
  \vskip 0.8truecm\noindent \vbox{\eightpoint\noindent
  {\bf\fig(#1)}: #4}\endinsert}}
\def\figurewithtex #1 #2 #3 #4 #5\cr{\null
  \global\advance\FIGUREcount by 1
  \NEWDEF fig,#1,{Fig.~\number\FIGUREcount}
  \write16{ FIG \number\FIGUREcount: #1}
  {\goodbreak\figcenter=\hsize\relax
  \advance\figcenter by -#4truecm
  \divide\figcenter by 2
  \midinsert\vskip #3truecm\noindent\hskip\figcenter
  \includegraphics{#1}{\hskip\texpscorrection\input #2 }
  \vskip 0.8truecm\noindent \vbox{\eightpoint\noindent
  {\bf\fig(#1)}: #5}\endinsert}}
\def\fig(#1){\ifundefined{fig#1}$\spadesuit$#1
  \else\csname fig#1\endcsname\fi}
%
%
\catcode`@=11
\def\footnote#1{\let\@sf\empty 
  \ifhmode\edef\@sf{\spacefactor\the\spacefactor}\/\fi
  #1\@sf\vfootnote{#1}}
\def\vfootnote#1{\insert\footins\bgroup\eightpoint
  \interlinepenalty\interfootnotelinepenalty
  \splittopskip\ht\strutbox 
  \splitmaxdepth\dp\strutbox \floatingpenalty\@MM
  \leftskip\z@skip \rightskip\z@skip \spaceskip\z@skip \xspaceskip\z@skip
  \textindent{#1}\footstrut\futurelet\next\fo@t}
\def\fo@t{\ifcat\bgroup\noexpand\next \let\next\f@@t
  \else\let\next\f@t\fi \next}
\def\f@@t{\bgroup\aftergroup\@foot\let\next}
\def\f@t#1{#1\@foot}
\def\@foot{\strut\egroup}
\def\footstrut{\vbox to\splittopskip{}}
\skip\footins=\bigskipamount 
\count\footins=1000 
\dimen\footins=8in  
\catcode`@=12       
%
\font\twelverm=cmr12
\font\twelvei=cmmi12
\font\twelvesy=cmsy10 scaled\magstep1
\font\twelveex=cmex10 scaled\magstep1
\font\twelvebf=cmbx12 
\font\twelvett=cmtt12
\font\twelvesl=cmsl12
\font\twelveit=cmti12
\font\ninerm=cmr9

\font\ninesy=cmsy9

\font\eightrm=cmr8
\font\eighti=cmmi8
\font\eightsy=cmsy8
\font\eightex=cmex8
\font\eightbf=cmbx8
\font\eighttt=cmtt8
\font\eightsl=cmsl8
\font\eightit=cmti8
\font\sixrm=cmr6
\font\sixi=cmmi6
\font\sixsy=cmsy6
\font\sixbf=cmbx6
\newfam\truecmr 
\newfam\truecmsy
\font\twelvetruecmr=cmr10 scaled\magstep1
\font\twelvetruecmsy=cmsy10 scaled\magstep1
\font\tentruecmr=cmr10
\font\tentruecmsy=cmsy10
\font\eighttruecmr=cmr8
\font\eighttruecmsy=cmsy8
\font\seventruecmr=cmr7
\font\seventruecmsy=cmsy7
\font\sixtruecmr=cmr6
\font\sixtruecmsy=cmsy6
\font\fivetruecmr=cmr5
\font\fivetruecmsy=cmsy5
\textfont\truecmr=\tentruecmr
\scriptfont\truecmr=\seventruecmr
\scriptscriptfont\truecmr=\fivetruecmr
\textfont\truecmsy=\tentruecmsy
\scriptfont\truecmsy=\seventruecmsy
\scriptscriptfont\truecmsy=\fivetruecmsy
%
\def \eightpoint{\def\rm{\fam0\eightrm}
  \textfont0=\eightrm \scriptfont0=\sixrm \scriptscriptfont0=\fiverm 
  \textfont1=\eighti  \scriptfont1=\sixi  \scriptscriptfont1=\fivei 
  \textfont2=\eightsy \scriptfont2=\sixsy \scriptscriptfont2=\fivesy 
  \textfont3=\eightex \scriptfont3=\eightex \scriptscriptfont3=\eightex
  \textfont\itfam=\eightit          \def\it{\fam\itfam\eightit}%
  \textfont\slfam=\eightsl          \def\sl{\fam\slfam\eightsl}%
  \textfont\ttfam=\eighttt          \def\tt{\fam\ttfam\eighttt}%
  \textfont\bffam=\eightbf          \scriptfont\bffam=\sixbf
  \scriptscriptfont\bffam=\fivebf   \def\bf{\fam\bffam\eightbf}%
  \textfont\truecmr=\eighttruecmr   \scriptfont\truecmr=\sixtruecmr
  \scriptscriptfont\truecmr=\fivetruecmr
  \textfont\truecmsy=\eighttruecmsy \scriptfont\truecmsy=\sixtruecmsy
  \scriptscriptfont\truecmsy=\fivetruecmsy
  \tt \ttglue=.5em plus.25em minus.15em 
  \setbox\strutbox=\hbox{\vrule height7pt depth2pt width0pt}%
  \normalbaselineskip=9pt
  \let\sc=\sixrm  \let\big=\eightbig  \normalbaselines\rm
}
\def \twelvepoint{\def\rm{\fam0\twelverm}
\textfont0=\twelverm  \scriptfont0=\tenrm  \scriptscriptfont0=\eightrm 
\textfont1=\twelvei   \scriptfont1=\teni   \scriptscriptfont1=\eighti 
\textfont2=\twelvesy  \scriptfont2=\tensy  \scriptscriptfont2=\eightsy 
\textfont3=\twelveex  \scriptfont3=\tenex  \scriptscriptfont3=\eightex 
\textfont\itfam=\twelveit             \def\it{\fam\itfam\twelveit}%
\textfont\slfam=\twelvesl             \def\sl{\fam\slfam\twelvesl}%
\textfont\ttfam=\twelvett             \def\tt{\fam\ttfam\twelvett}%
\textfont\bffam=\twelvebf             \scriptfont\bffam=\tenbf
\scriptscriptfont\bffam=\eightbf      \def\bf{\fam\bffam\twelvebf}%
\textfont\truecmr=\twelvetruecmr      \scriptfont\truecmr=\tentruecmr
\scriptscriptfont\truecmr=\eighttruecmr
\textfont\truecmsy=\twelvetruecmsy    \scriptfont\truecmsy=\tentruecmsy
\scriptscriptfont\truecmsy=\eighttruecmsy
\tt \ttglue=.5em plus.25em minus.15em 
\setbox\strutbox=\hbox{\vrule htwelve7pt depth2pt width0pt}%
\normalbaselineskip=12pt
\let\sc=\tenrm  \let\big=\twelvebig  \normalbaselines\rm
}
\catcode`@=11
\def\eightbig#1{{\hbox{$\textfont0=\ninerm\textfont2=\ninesy\left#1
  \vbox to6.5pt{}\right.\n@space$}}}
\catcode`@=12


%% file: defs.tex
%

\def\CC{{\cal C}}

\def\EE{{\cal E}}
\def\FF{{\cal F}}

\def\LL{{\cal L}}
\def\MM{{\cal M}}
\def\NN{{\cal N}}
\def\OO{{\cal O}}

\def\HB {\hfill\break}
\def\sqr#1#2{{\vcenter{\vbox{\hrule height .#2pt
        \hbox{\vrule width.#2pt height#1pt \kern#1pt
           \vrule width.#2pt}
        \hrule height.#2pt}}}}
\def\square{\mathchoice\sqr64\sqr64\sqr{2.1}3\sqr{1.5}3}        
\def\QED{\null\hfill$\square$}

\def\eqdef{\buildrel\hbox{\sevenrm def}\over =}
\def\real{{\bf R}}
\def\natural{{\bf N}}
\def\complex{{\bf C}}

\def\Re{{\rm Re\,}}

\def\L{{\rm L}}
\def\H{{\rm H}}
\def\Z{{\rm Z}}
\def\X{{\rm X}}

\def\epsilon{\varepsilon}
\def\phi{\varphi}

\def\ttau{\hat \tau}
\def\ie{{\sl i.e.\ }}
\def\x{{\rm x}}
\def\t{{\rm t}}
\def\C{\tilde C}

\def\e{{\rm e}}

\def\d{{\rm \,d}}
\def\dd{{\rm d}}
\def\frac#1#2{{\textstyle{#1 \over #2}}}

%% file: refs.tex
\Item{AW}{AW} D.G. Aronson and H.F. Weinberger: 
  {\pap Multidimensional Nonlinear Diffusion Arising in Population Genetics},
  Adv. in Math. {\bf 30} (1978), 33--76.   

\Item{Br}{Br} M. Bramson:
  {\bok Convergence of Solutions of the Kolmogorov Equation to Travelling
  Waves}, Memoirs of the AMS {\bf 44}, nb. 285, Providence (1983). 

\Item{BK1}{BK1} J. Bricmont and A. Kupiainen: {\pap Stability of Moving
  Fronts in the Ginzburg-Landau Equation}, Commun.\ Math.\ Physics {\bf 159} 
  (1994), 287--318. 

\Item{BK2}{BK2} J.  Bricmont and A.  Kupiainen: {\pap Stable Non-Gaussian
  Diffusive Profiles}, Nonlinear Anal., Theory Methods Appl. {\bf 26}
  (1996), 583--593.

\Item{CH}{CH} Th.  Cazenave and A.  Haraux: {\bok Introduction aux
  Probl\`emes d'Evolution Semi-lin\'eaires}, Math\'ematiques et
  Applications {\bf 1}, Ellipses (1990).

\Item{DO}{DO} S.R. Dunbar and H.G. Othmer: {\pap On a Nonlinear Hyperbolic
  Equation Describing Transmission Lines, Cell Movement, and Branching
  Random Walks}, in {\bok Nonlinear Oscillations in Biology and Chemistry}, 
  H.G. Othmer (Ed.), Lect. Notes in Biomathematics {\bf 66}, Springer (1986). 

\Item{EW}{EW} J.-P. Eckmann and C.E. Wayne:
  {\pap The Nonlinear Stability of Front Solutions for Parabolic Partial
  Differential Equations}, Comm. Math. Phys. {\bf 161} (1994), 323--334. 

\Item{EKM}{EKM} M. Escobedo, O. Kavian and H. Matano: {\pap Large Time
  Behavior of Solutions of a Dissipative Semi-linear Heat Equation},
  Comm.  Partial Diff.  Equations {\bf 20} (1995), 1427--1452.

\Item{EZ}{EZ} M. Escobedo and E. Zuazua: {\pap Large-time Behavior
  for Convection Diffusion Equations in $\real^N$}, J.  Funct.  Anal.
  {\bf 100} (1991), 119--161.

\Item{Fi}{Fi} R.A. Fisher:
  {\pap The Advance of Advantageous Genes}, Ann. of Eugenics {\bf 7} (1937), 
  355--369. 

\Item{GV}{GV} V.A. Galaktionov and J. L. Vazquez:
  {\pap Asymptotic Behaviour of Nonlinear Para\-bolic Equations with Critical 
  Exponents. A Dynamical System Approach}, J. Funct. Anal. {\bf 100} (1991), 
  435--462.

\Item{Ga}{Ga} Th. Gallay:
  {\pap Local Stability of Critical Fronts in Nonlinear Parabolic
  Partial Differential Equations}, Nonlinearity {\bf 7} (1994), 741--764. 

\Item{GM}{GM} Th. Gallay and A. Mielke: {\pap Diffusive Mixing of Stable
  States in the Ginzburg-Landau Equation}, Commun. Math. Phys. {\bf 199}
  (1998), 71--97. 

\Item{GR1}{GR1} Th.  Gallay and G.  Raugel: {\pap Stability of Travelling
  Waves for a Damped Hyperbolic Equation}, ZAMP {\bf 48} (1997),
  451--479.

\Item{GR2}{GR2} Th.  Gallay and G.  Raugel: {\pap Scaling Variables and 
  Asymptotic Expansions in Damped Wave Equations}, to appear in 
  J. Diff. Eqns (1998). 

\Item{GR3}{GR3} Th.  Gallay and G.  Raugel: {\pap Stability of Propagating
  Fronts in Damped Hyperbolic Equations}, in: Proceedings of the conference 
  ``Partial differential equations: theory and numerical solutions'' held 
  in Prague (1998), Longman, to appear. 

\Item{Go}{Go} S. Goldstein: {\pap On Diffusion by Discontinuous Movements
  and the Telegraph Equation}, Quart. J. Mech. Appl. Math. {\bf 4}
  (1951), 129--156.

\Item{Ha1}{Ha1} K.P. Hadeler:
  {\pap Hyperbolic Travelling Fronts}, Proc. Edinb. Math. Soc. {\bf 31} 
  (1988), 89--97.   

\Item{Ha2}{Ha2} K.P. Hadeler:
  {\pap Reaction Telegraph Equations and Random Walk Systems}, 
  in: {\bok Stochastic and spatial structures of dynamical systems},
  S. van Strien, S. Verduyn Lunel (eds.), Royal Acad. of the Netherlands,
  North Holland, Amsterdam (1996).  

\Item{Ha3}{Ha3} K.P. Hadeler:
  {\pap Reaction Transport Systems}, in: {\bok Mathematics
  inspired by biology}, V.Capasso, O.Diekmann (eds), CIME Lectures 1997, 
  Florence, Springer Verlag, in print.

\Item{HLP}{HLP} G.H.  Hardy, J.E.  Littlewood and G.  Polya: {\bok
  Inequalities}, Cambridge University Press (1934), reprinted (1964).

\Item{HL}{HL} L. Hsiao and T.-P. Liu: {\pap Convergence to Nonlinear
  Diffusion Waves for Solution of a System of Hyperbolic Conservation
  Laws with Damping}, Comm.  Math.  Phys.  {\bf 143} (1992), 599--605.

\Item{Kac}{Kac} M. Kac: {\pap A Stochastic Model Related to the
  Telegrapher's Equation}, Rocky Mountain J. Math. {\bf 4} (1974),
  497--509.

\Item{Kap}{Kap} T. Kapitula:
  {\pap On the Stability of Travelling Waves in Weighted $L^\infty$ 
  Spaces}, J. Diff. Eqns {\bf 112} (1994), 179--215. 

\Item{Kav}{Kav} O. Kavian: {\pap Remarks on the Large Time Behavior of a
  Nonlinear Diffusion Equation}, Ann.  Inst.  Henri Poincar\'e {\bf 4}
  (1987), 423--452.

\Item{Ki}{Ki} K. Kirchg\"assner:
  {\pap On the Nonlinear Dynamics of Travelling Fronts}, J. Diff. Eqns. 
  {\bf 96} (1992), 256--278. 

\Item{KPP}{KPP} A.N. Kolmogorov, I.G. Petrovskii and N.S. Piskunov:
  {\pap Etude de la diffusion avec croissance de la quantit\'e de mati\`ere
  et son application \`a un probl\`eme biologique}, Moscow Univ. Math. Bull.
  {\bf 1} (1937), 1--25. 

\Item{Mu}{Mu} J.D. Murray: {\bok Mathematical Biology} 2nd ed.,
  Biomathematics {\bf 19}, Springer Verlag (1993).

\Item{Ni}{Ni} K. Nishihara: {\pap Convergence Rates to Nonlinear
  Diffusion Waves for Solutions of System of Hyperbolic Conservation
  Laws with Damping}, J.  Differential Equations {\bf 131} (1996),
  171--188.

\Item{RK}{RK} G. Raugel, K. Kirchg\"assner: {\pap Stability of Fronts
  for a KPP-System, II: The Critical Case}, J.  Differential Equations 
  {\bf 146} (1998), 399--456. 

\Item{Sa}{Sa} D.H. Sattinger:
  {\pap On the Stability of Waves of Nonlinear Parabolic Systems}, 
  Adv. Math. {\bf 22} (1976), 312--355. 

\Item{Wa}{Wa} C.E. Wayne: {\pap Invariant Manifolds for Parabolic
  Partial Differential Equations in Unbounded Domains}, Arch. Rat.
  Mech. Anal. {\bf 138} (1997), 279--306. 

\Item{Za}{Za} E. Zauderer: {\bok Partial Differential Equations of
  Applied Mathematics}, New York, John Wiley (1983).